\documentclass[proof]{WileyASNA-v1}

\usepackage{color}

\articletype{Article Type}%

\received{23 December 2020}
\revised{28 January 2021}
\accepted{29 January 2021}

\begin{document}

\title{Rubidium abundances of galactic disk stars}

\author[]{Yoichi Takeda}

\authormark{Y. TAKEDA}

\address[]{ 
\orgaddress{\state{11-2 Enomachi, Naka-ku, Hiroshima-shi, 730-0851}, \country{Japan}}}

\corres{\email{ytakeda@js2.so-net.ne.jp}}


\abstract{
Spectroscopic determinations of Rubidium abundances were conducted by applying the spectrum 
fitting method to the Rb~{\sc i} 7800 line for an extensive sample of $\sim 500$ late-type 
dwarfs as well as giants (including Hyades cluster stars) belonging to the galactic disk 
population, with an aim of establishing the behaviour of [Rb/Fe] ratio for disk stars 
in the metallicity range of $-0.6 \lesssim$~[Fe/H]~$\lesssim +0.3$.
An inspection of the resulting Rb abundances for Hyades dwarfs revealed that they show 
a systematic $T_{\rm eff}$-dependent trend at 
$\gtrsim 5500$~K; this means that the results for mid-G to F stars (including the Sun) are 
not reliable (i.e., more or less overestimated), which might be due to some imperfect 
treatment of surface convection in classical model atmospheres.
As such, it was decided to confine only to late-G and K stars at $T_{\rm eff} \lesssim 5500$~K
and adopt the solar-system (meteoritic) value as the reference Rb abundance.
The [Rb/Fe] vs.[Fe/H] relations derived for field dwarfs and giants turned out to be consistent 
with each other, showing a gradual increase of [Rb/Fe] with a decrease in [Fe/H]
(with d[Rb/Fe]/d[Fe/H] gradient of $\sim -0.4$ around the solar metallicity),
which is favourably compared with the theoretical prediction of chemical evolution models.
Accordingly, this study could not confirm the anomalous behaviour of [Rb/Fe] ratio (tending to 
be subsolar but steeply increasing toward supersolar metallicity) recently reported 
for M dwarf stars of $-0.3 \lesssim$~[Fe/H]~$\lesssim +0.3$. 
}

\keywords{Galaxy: disk --- Galaxy: evolution --- stars: abundances --- 
stars: atmospheres ---  stars: late-type
}

\maketitle

\footnotetext{\textbf{Abbreviations:} LTE, local thermodynamic equilibrium; 
LIMS, low- and intermediate-mass stars; MSR, rotating massive star}

\section{Introduction}
 
Rubidium (Rb, $Z = 37$) is a neutron-capture element (in which both s- and r-processes 
are involved), whose abundances spectroscopically determined for various types of stars 
may provide us with useful information regarding mixing of nuclear-process products 
or galactic chemical evolution.
However, published spectroscopic studies of stellar Rb abundances have been rather limited
in number, which may presumably reflect the comparatively large technical difficulty 
of its abundance determination (i.e., weak and blended line feature has to be dealt
in many cases).

Those investigations on Rb abundances so far have focused mainly on metal-poor stars.
Following Gratton \& Sneden (1994) who determined the abundances of Rb (along with other
neutron-rich elements) for 19 stars at $-2.8 <$~[Fe/H]~$< 0$,
Tomkin \& Lambert (1999) conducted a more detailed Rb abundance analysis for 44 
metal-deficient giants and dwarfs in the metallicity range of $-2.0<$~[Fe/H]~$< 0.0$.
These previous studies revealed that the [Rb/Fe] ratio tends to be moderately supersolar  
($0 \lesssim$~[Rb/Fe]~$\lesssim 0.5$) at the metal-poor regime ([Fe/H]~$\lesssim -1$).
Successively, Rb abundances of globular cluster giants (Yong et al. 2006, 2008; 
D'Orazi et al. 2013) or AGB giants (Garc\'{\i}a-Hern\'{a}ndez et al. 2006;
Shejeelammal et al. 2020) were also investigated.

On the other hand, in contrast to such metal-poor halo stars or peculiar AGB stars, 
the behaviours of Rb abundances in galactic disk stars (mildly metal-deficient through 
metal-rich regime) seem to have been poorly investigated. What can be read from 
a small number of samples at $-0.5 \lesssim$~[Fe/H]~$\lesssim 0.0$ in previous studies 
quoted above is that [Rb/Fe] tends to be slightly subsolar by 0.1--0.3~dex 
(see, e.g., Fig.~2 in Gratton \& Sneden 1994 or Fig.~3 in Tomkin \& Lambert 1999).

In this context, the problem of reference Rb abundance should be mentioned, which
might have something to do with this trend. That is,  Gratton \& Sneden (1994)
as well as Tomkin \& Lambert (1999) similarly remarked that 
the solar Rb abundance\footnote{Throughout this paper, $A$ denotes
the logarithmic number abundance in the usual normalisation of $A$(H) = 12.00.}
 $A_{\odot} \simeq 2.6$, which they determined from the solar spectrum and used as
the fiducial abundance for evaluation of [Rb/Fe]
($\equiv A_{*} - A_{\odot} -$~[Fe/H]), is larger than the solar-system 
Rb abundance ($A_{\rm s.s.} \simeq 2.4$) well established from meteorites,
though they could not find out the cause for this difference.  
If the latter $A_{\rm s.s.}$ were used as the zero-point abundance, the subsolar 
tendency would be mitigated in the direction of [Rb/Fe]$\sim 0$.   

Recently, an interesting finding has been reported by Abia et al. (2020), 
who examined the Rb abundances of 57 M dwarfs in the metallicity
range $-0.3 \lesssim$~[Fe/H]~$\lesssim +0.3$ observed with CARMENES.
The result they obtained was somewhat surprising:  The [Rb/Fe] ratios
of many stars turned out to be subsolar by a few tenths dex and show an increasing 
tendency towards higher metallicity from [Rb/Fe]~$\sim -0.4$ (at [Fe/H]~$\sim -0.3$)
to [Rb/Fe]~$\sim +0.2$ (at [Fe/H]~$\sim +0.3$). Such a trend is unusual for 
a neutron-capture element and can not be explained even by the state-of-the-art chemical 
evolution models. So, if this is real, a significant revision would be required
on the theoretical side. Yet, given that reliability of abundance determinations 
in M dwarfs has not been firmly established, follow-up verification should be
necessary. Abia et al. (2020) themselves remarked ``{\it Additional Rb abundance
measurements in FGK dwarfs of near solar metallicity, as well as an evaluation
of the impact of stellar activity on abundance determinations in M dwarfs,
are urgently needed to confirm or disprove the main findings of this study.}''

This problem attracted the author's interest and eventually motivated this study.
Conveniently, observational data for Rb abundance determination are already 
available for a large number of FGK-type stars belonging to the disk population 
(field dwarfs as well as giants, Hyades cluster stars) in the relevant metallicity 
range ($-0.6 \lesssim$~[Fe/H]~$\lesssim +0.3$), for which atmospheric parameters 
were also established in the previous papers. 
Accordingly, it was decided to determine the Rb abundances for these sample stars  
and check up their behaviours, while paying attention also to examining the credibility
of resulting abundances. Specifically, the purpose of this investigation was 
to clarify the following points: 
\begin{itemize}
\item
Do the Rb abundances resulting from various Hyades stars of F-, G-, and K-type agree  
with each other? This would serve as a reliability check for the adopted procedure of 
abundance determination. 
\item
Are the Rb abundances derived from dwarf and giant stars consistent with each other
at the same [Fe/H]? 
\item
Is the solar photospheric Rb abundance determined from the spectrum of the Sun (or Moon)
agree with the solar-system (meteorite) abundance? Or does it show an appreciable 
discrepancy such as found in previous studies?
\item
Finally, how do the [Rb/Fe] ratios of disk stars (especially those of near-solar metallicity) 
behave itself with a change in [Fe/H]? Could the trend reported by Abia et al. (2020) be confirmed?
\end{itemize}

\section{Program stars and their parameters}

A total of 487 stars were selected from the samples of previous studies 
of the author, which are divided into 4 groups (group~1 --- 47 Hyades FGK dwarfs, 
group~2 --- 100 field FGK dwarfs, group~3 --- 101 field GK dwarfs, 
and group~4 --- 239 field GK giants) as summarised in Table~1, where the references 
to the original papers are also given for more detailed information.

Regarding the basic observational data, the spectra obtained by using HIDES (HIgh-Dispersion 
Echelle Spectrograph) attached to the 188~cm reflector at Okayama Astrophysical Observatory  
were used in most cases, though those acquired by HDS (High-Dispersion Spectrograph) with
8.2~m Subaru Telescope were partly used for 13 (out of 47) group~1 stars
and 17 (out of 101) group~3 stars.

As to the reference solar spectrum, the Moon spectrum\footnote{Although this spectrum is 
damaged in the short-wavelength side of the important Rb~{\sc i} 7800 line like 
other spectra of many group~2 stars taken from Takeda et al.'s (2005a) database 
(cf. the note to Table~1), the whole profile of Si~{\sc i}+Rb~{\sc i} line feature 
turned out fortunately usable (cf. Fig.~1).} included in Okayama spectrum database 
(Takeda et al. 2005a) was employed 
along with Neckel's (1994, 1999) solar disk-centre spectrum and Kurucz et al.'s (1984) 
solar flux spectrum.

The atmospheric parameters [$T_{\rm eff}$ (effective temperature), $\log g$ (surface gravity),
$v_{\rm t}$ (microturbulence), and [Fe/H] (metallicity)] of the program stars were 
spectroscopically determined in the same consistent manner from the equivalent widths 
of Fe~{\sc i} and Fe~{\sc ii} lines by using the TGVIT program (Takeda et al. 2005b). 
These stellar parameters are given in the online materials 
(tableE1,dat, tableE2.dat, tableE3.dat, and tableE4.dat) for each of the 4 groups. 
As seen from the parameter ranges for each group indicated in Table~1,
while the temperature ranges of group~1 and group~2 stars widely extend
from F- through K-type, those of group~3/group~4 are limited to a rather narrow span of 
$\sim 1000$~K (i,e., $\sim 5500$--4500~K corresponding to late/mid~G--mid/early~K).

Similarly, the same atmospheric models were used as employed in the original references;
they are based on Kurucz's (1993a) ATLAS9 model atmosphere grids (computed with 
the convective overshooting effect), which were interpolated in terms of 
$T_{\rm eff}$, $\log g$, and [Fe/H] for each star.   

\section{Abundance determination}

\subsection{Outline of the method}

Regarding determinations of Rb abundances and related quantities, the resonance Rb~{\sc i} 
line at 7800~\AA\ (which comprises multiple components and is blended with the Si~{\sc i} line) 
is exclusively used in this study.\footnote{The other resonance line of the doublet
at 7947~\AA\ was not used, because it is weaker and the quality of observed spectra
in this wavelength region is poor (e.g., contamination of telluric lines) compared
to the Rb~{\sc i} 7800 line.} 
The procedures are essentially the same as those adopted in the author's previous papers 
(e.g., Takeda \& Kawanomoto 2005, Takeda et al. 2015, Takeda \& Honda 2020). 
First, by applying Takeda's (1995) numerical algorithm, the best fit between the synthetic 
and observed spectra is accomplished in the specified wavelength region (7776--7804~\AA) 
while varying the abundances of important elements ($A$(Ni), $A$(Si), $A$(Rb) and 
$A$(Fe) in the present case), $v_{\rm M}$ (macrobroadening velocity\footnote{This $v_{\rm M}$ 
(including instrumental broadening, rotational broadening, and macroturbulence) is the 
$e$-folding half-width of the Gaussian broadening function $f(v) \propto \exp[-(v/v_{\rm M})^{2}]$.}), 
and $\Delta \lambda$ (radial velocity or wavelength shift). 
Then, the equivalent width ($W$)\footnote{It should be noted that this $W$ is evaluated by integrating 
the synthesised line profile on the assumption that this Rb~{\sc i}~7800 line feature (comprising 
multiple components as listed in Table~2) is isolated, despite that it is actually blended with 
the neighbouring Si~{\sc i} line. Therefore, it is not so much a directly observable 
quantity in real spectra as rather a useful measure of line strength.} 
of the relevant Rb~{\sc i} 7800 line is ``inversely'' evaluated from the best-fit solution 
of $A$(Rb) with the same atmospheric model/parameters as used in the spectrum-fitting analysis. 
From such evaluated $W$, the non-LTE abundance ($A^{\rm N}$) and LTE abundance ($A^{\rm L}$) 
are calculated, from which the non-LTE correction is derived as $\Delta \equiv A^{\rm N} - A^{\rm L}$.  
Besides, this $W$ can be further used to estimate the abundance uncertainties 
due to ambiguities of atmospheric parameters by perturbing the standard values interchangeably. 

\subsection{Adopted data of spectral lines}

Regarding the atomic data (wavelengths, oscillator strengths) of the Rb~{\sc i} 7800 line, 
Abia et al.'s (2020) Table~1 was invoked. The $gf$ values of the component lines were scaled 
by assuming the solar-system (meteoritic) isotope ratio of $^{85}$Rb/$^{87}$Rb~=~2.43
as done by them. These scaled data are listed in Table~2.
As for the damping parameters, the default treatment of Kurucz's (1993a) WIDTH9 program was
applied, though they are essentially insignificant in the present case because this 
Rb~{\sc i} line is generally weak. 
Otherwise, the data compiled in the VALD database (Ryabchikova et al. 2015) were used 
for the lines (other than Rb~{\sc i} 7800) contained in the 7776--7804~\AA\ region.
The exceptions were the following three Si~{\sc i} lines ($\chi_{\rm low} = 6.181$~eV) 
whose $gf$ values were empirically adjusted: 
Si~{\sc i} 7798.837 ($\log gf = -2.06$),
Si~{\sc i} 7799.180 ($\log gf = -1.75$), and
Si~{\sc i} 7799.996 ($\log gf = -0.69$).
The abundances of all elements other than Ni, Si, Rb, and Fe were fixed at the metallicity-scaled
solar abundances in the spectrum-fitting analysis. 

\subsection{Non-LTE calculations}

The statistical-equilibrium calculations for neutral rubidium
were carried out in the similar manner as previously done for the lighter 
alkali elements: Li (Takeda \& Kawanomoto 2005), K (Takeda et al. 1996; 2002),
and Na (Takeda et al. 2003). For the present study, Rb~{\sc i} model atom 
was constructed comprising 19 terms (up to 11s~$^{2}$S at 31917 cm$^{-1}$)
and 59 radiative transitions, while using Kurucz \& Bell's (1995) compilation 
of atomic data ($gf$ values, levels, etc.). 

Regarding the treatment of the photoionisation cross section, the hydrogenic 
approximation was assumed, except for the edge values of the ground term (5s~$^{2}$S) 
and the first excited term (5p~$^{2}$P$^{\circ}$), for which  Lowell et al. (2002) 
and Nadeem \& Haq (2011) were invoked, respectively.  
As to the collisional rates, the recipe described in Sect.~3.1.3 of Takeda (1991)
was basically followed, while the H~{\sc i} collision rates computed with 
the classical Drawin's cross section (cf. Steenbock \& Holweger 1984) was multiplied 
by a factor of $k = 10^{-3}$ (i.e., suppressed to a negligible level).\footnote{
This $k$ value was adopted by consulting the previous cases for the resonance 
lines of other alkali species; i.e., $k = 10^{-3}$ for Li~{\sc i} 6708 as well as 
K~{\sc i} 7699, $k = 10^{-1}$ for Na~{\sc i} 5890/5896.
Actually, however, the choice of this correction factor does not matter much, because
the resulting non-LTE effect turns out anyhow insignificant even by drastically 
reducing the H~{\sc i} collisions as such. For example, the $|\Delta|$ values are 
$\lesssim 0.1$~dex at most (cf. Sect.~3.4), which are further reduced by $\lesssim$ several 
hundredths dex if the H~{\sc i} collision effect is included with $k=1$.}    

Since FGK-type stars (dwarfs as well as giants) of population~I in the galactic 
disk are concerned in this study, the calculations were done on a grid of 
125 ($5 \times 5 \times 5$) model atmospheres resulting from 
combinations of five $T_{\rm eff}$ values 
(4500, 5000, 5500, 6000, 6500~K), five $\log g$ values 
(1.0, 2.0, 3.0, 4.0, 5.0), and five [Fe/H] values 
(+0.5, 0.0, $-0.5$, $-1.0$, $-1.5$).
The microturbulent velocity was assumed to be $v_{\rm t}$ = 2~km~s$^{-1}$,
and the metallicity-scaled Rb abundance ([Rb/Fe] = 0) was used as 
$A$(Rb) = 2.60 + [Fe/H], where 2.60 is Anders \& Grevesse's (1989) 
solar Rb abundance adopted in ATLAS9 models. 
The depth-dependent non-LTE departure coefficients to be used for each star were then
evaluated by interpolating this grid in terms of $T_{\rm eff}$, $\log g$, and [Fe/H].

\subsection{Results}

The Rb abundances were successfully determined for almost all cases, except for
3 stars in group~2 (HD~120136, 124850, 016673; these are F-type dwarfs with
$T_{\rm eff} \gtrsim 6300$~K and the Rb~{\sc i} 7800 line is very weak), for which
$A$(Rb) solution was not converged and thus had to be fixed in the fitting process.
Representative examples of spectrum fitting in the whole 7796--7804~\AA\ region 
are shown in Fig.~1 (for the disk-centre Sun, disk-integrated Sun, 
Moon, and 8 dwarfs/giants in the wide $T_{\rm eff}$ range).
The close-up display of the accomplished fit in the 7799--7801~\AA\ region is
presented for all program stars in Fig.~2 (groups 1, 2, and 3) and Fig.~3 (group 4).

After the $A$(Rb) solution has been established, the related quantities were further
derived as described in Sect.~3.1: $W$ (equivalent width), $\Delta$ (non-LTE 
correction), $A^{\rm N}$ (non-LTE abundance), $\delta_{T\pm}$ (abundance changes for 
$T_{\rm eff}$ perturbations by $\pm 100$~K), $\delta_{g\pm}$ (abundance changes for 
$\log g$ perturbations by $\pm 0.1$~dex), and $\delta_{v\pm}$ (abundance changes for 
$v_{\rm t}$ perturbations by $\pm 0.5$~km~s$^{-1}$). Each of them are plotted against 
$T_{\rm eff}$ in Fig~4 (group 1), Fig.~5 (group 2), Fig.~6 (group 3), and Fig.~7 (group 4).
These results of $W$, $\Delta$, $A^{\rm N}$  (along with $v_{\rm M}$ derived from 
spectrum fitting as by-product) for stars of each group are presented in 
the online tables (tableE1.dat, tableE2.dat, tableE3.dat, and tableE4.dat). 

An inspection of Fig.~4--7 suggests the following trends which generally hold for
all program stars irrespective of dwarfs or giants.
\begin{itemize}
\item
The equivalent width of Rb~{\sc i} 7800 ($W$) decreases with an increase 
in $T_{\rm eff}$, reflecting that the occupation number of the Rb~{\sc i} ground 
level is quite $T_{\rm eff}$-sensitive ($\propto 10^{+5040 \chi^{\rm ION}/T_{\rm eff}}$ 
where $\chi^{\rm ION}$ = 4.18~eV is the ionisation potential of Rb~{\sc i}).
\item
The non-LTE correction ($\Delta$) progressively increases as $T_{\rm eff}$ is lowered: 
While it is negative ($\Delta \sim -0.1$~dex: non-LTE line strengthening) for F-type stars, 
it becomes positive ($\Delta \sim +0.1$~dex: non-LTE line weakening) for K-type stars.
As such, the non-LTE effect on Rb abundance determination ($|\Delta| \lesssim 0.1$~dex) is 
insignificant as far as stars at 4500~$\lesssim T_{\rm eff} \lesssim 6500$~K are concerned.
\item
Among the various sources of abundance errors, most important is $\delta_{T\pm}$   
($\sim \pm$~0.1~dex for $\Delta T_{\rm eff} = \pm 100$~K; tending to enhance toward
lower $T_{\rm eff}$), which can be naturally understood from the large $T_{\rm eff}$-sensitivity
of $W$ mentioned above. Meanwhile, $\delta_{g\pm}$ and $\delta_{v\pm}$ are totally insignificant.
\end{itemize} 

The Rb abundances determined in this investigation are compared with those derived by 
Tomkin \& Lambert (1999) in Fig.~8 for 6 stars in common, where a satisfactory match 
is observed for most cases. 
One exception showing a rather large deviation ($\sim$~0.3--0.4~dex) is the G5V star 
$\mu$~Cas (= HD~6582), for which Rb abundance determination is considerably difficult 
because the Rb line is very weak ($W = 1.6$~m\AA) due to its low metallicity 
([Fe/H] = $-0.8$). 

\section{Discussion}

\subsection{Solar Rb abundance problem}

Let us first discuss the solar rubidium abundance to be used as the reference standard,
for which appreciable discrepancy from the meteoritic composition was reported in
previous work (cf. Sect.~1).
The non-LTE Rb abundance determined from Kurucz et al.'s (1984) solar flux
spectrum is $A^{\rm N}_{\odot} = 2.44$ ($W = 5.9$~m\AA, $\Delta = -0.05$~dex),
that from Neckel's (1994, 1999) solar disk-centre spectrum (direction cosine $\mu = \cos\theta = 1$)
is $A^{\rm N}_{\odot} = 2.48$ ($W^{\mu=1} = 5.7$~m\AA, $\Delta = -0.05$~dex), 
while that derived from Takeda et al.'s (2005) Moon spectrum is
$A^{\rm N}_{\odot} = 2.56$ ($W = 7.7$~m\AA, $\Delta = -0.06$~dex),
resulting in a Sun$-$Moon difference of $\sim$~0.1~dex despite that a satisfactory 
fit appears to be accomplished similarly for both cases (cf. Fig.~1).
This suggests that Rb abundance determination from a severely blended feature 
is a delicate matter influenced by the difference in the fitting region
(which had to be narrowed in the latter case owing to the spectrum defect 
in the short-wavelength side) or in the spectrum resolving power.
Accordingly, these spectroscopically established solar Rb abundances are 
higher than the solar-system abundance ($A_{\rm s.s.} = 2.36$; cf. Lodders 2020) 
by $\sim$~0.1--0.2~dex.
As such, the discrepancy between $A_{\odot}$ and $A_{\rm s.s.}$ reported by 
Gratton \& Sneden (1994) and Tomkin \& Lambert (1999) has been confirmed 
in this study.

In contrast, Abia et al. (2020) and Korotin (2020) claimed that this problem has been settled, 
since they obtained $A^{\rm N}_{\odot} = 2.35 \pm 0.05$ which is in very good agreement
with $A_{\rm s.s.}$ of 2.36. This is mainly because they adopted a more enhanced non-LTE 
correction ($\Delta = -0.12$~dex) than the value ($-0.05$~dex) mentioned above,
which reflects the fact that their non-LTE abundance corrections for FGK-type stars 
are systematically shifted downward by $\sim 0.1$~dex compared to those derived 
in this study (see Appendix~A where this difference is discussed more in detail). 
Yet, it is not clear whether this solar abundance issue is simply attributed to the non-LTE 
effect, because the situation appears to be more complex and some other $T_{\rm eff}$-dependent 
problem may be involved, as described in the next two sections. 

\subsection{Implication from Hyades dwarfs}

Comparing the Rb abundances of group~1 stars (47 FGK dwarfs belonging to the Hyades cluster  
which distribute at 4500~K~$\lesssim T_{\rm eff} \lesssim 6300$~K) with each other makes 
a valuable test for checking the validity of abundance determination, because similar 
abundances should be observed for such cluster stars born from the gas of same 
chemical composition. 
An inspection of Fig.~4c suggests an interesting trend that (i) $A^{\rm N}$ values for
Hyades dwarfs at $T_{\rm eff} \lesssim 5500$~K (mid-G through K type) are almost similar
irrespective of $T_{\rm eff}$, while (ii) those at $T_{\rm eff} \gtrsim 5500$~K (F through 
early G type) show an increasing tendency towards higher $T_{\rm eff}$.
  
Regarding the former lower $T_{\rm eff}$ group, it is reasonable to regard that 
the abundances are reliably established. Actually, the mean abundance  
($\langle A^{\rm N} \rangle =  2.56 \pm 0.10$) derived for 18 stars of $T_{\rm eff} < 5500$~K
is consistent with the expected primordial Rb abundance of this cluster ($A_{\rm s.s.}$ + 
[Fe/H]$_{\rm Hyades}$), where $A_{\rm s.s.}$ is the solar system abundance (2.36) and 
[Fe/H]$_{\rm Hyades}$ is the Hyades metallicity ($\simeq +0.2$; cf. Takeda \& Honda 2020).

As to the latter higher $T_{\rm eff}$ group, it should be remarked that errors involved with 
the resulting abundances must be larger, especially for F-type stars ($T_{\rm eff} \gtrsim 6000$~K), 
because the line feature becomes considerably weak (due to higher $T_{\rm eff}$) and broad/shallow 
(due to larger rotational velocity).\footnote{Note that the error bars attached to the data points
in Fig.~4c represent only the errors due to ambiguities of atmospheric parameters
and thus do not include those due to photometric noises or due to any systematic effect
caused by imperfect fitting.
The uncertainty of $W$ ($\delta W$) due to random noise was evaluated by using the relation derived by 
Cayrel (1988) (as done in  Sect.~4.2 of Takeda \& Honda 2020), which depends on S/N, line width, 
and pixel size. For a typical S/N  of $\sim 100$,  $\delta W$ turned out to be $\sim$~1--2~m\AA\
in most cases ($\lesssim 3$~m\AA\ for the broad-line F-type stars of $v\sin i \sim$~10--20~km~s$^{-1}$).
Accordingly, abundance results derived from very weak lines (i.e., $W$ less than several 
m\AA) significantly suffer from this error and thus should not be seriously taken. 
Otherwise, the effect of $\delta W$ on the abundance is practically insignificant. 
} Accordingly, those data points showing prominent deviations
(e.g., $A^{\rm N} \gtrsim 3$ or $A^{\rm N} \lesssim 2$) should not be seriously taken.
Even so, a global trend is recognised from Fig.~4c that $A^{\rm N}$ gradually increases
from $\sim 2.6$ (at $T_{\rm eff} \sim 5500$~K) to $\sim 2.9$ (at $T_{\rm eff} \sim 6300$~K),
which means that Rb abundances in this $T_{\rm eff}$ range suffer some $T_{\rm eff}$-dependent
systematic errors (i.e., overestimation with its extent growing towards higher $T_{\rm eff}$).  

\subsection{Abundance distribution of near-solar metallicity stars}

In order to check on this problem further for other field stars, stars with metallicities 
around the solar value ($-0.3 <$~[Fe/H]~$< +0.3$) were sorted out from each group (2, 3, and 4)
and the distributions of their Rb abundances were examined. The resulting $A^{\rm N}$ vs. 
$T_{\rm eff}$ plots and the histograms of $A^{\rm N}$ are displayed in Fig.~9, where the subset of 
especially close-to-solar-metallicity stars ($-0.1 <$~[Fe/H]~$< +0.1$) are separately shown 
with the pink colour.
Let us pay attention to Fig.~9a, which shows that the Rb abundances derived for group~2 stars 
of near-solar metallicity (most of them are F--G dwarfs of $T_{\rm eff} \gtrsim 5500$~K) 
tend to systematically increase with $T_{\rm eff}$, just like the case of Hyades dwarfs 
discussed in the previous section. It  has thus been corroborated that the Rb abundances 
of $T_{\rm eff} \gtrsim 5500$~K dwarfs are not so trustable in the sense that they tend to be 
overestimated (this effect begins at $\sim 5500$~K and grows with increasing $T_{\rm eff}$).

The cause of this embarrassing effect is not clear, which is limited only to stars of 
$T_{\rm eff} \gtrsim 5500$~K (while the lower $T_{\rm eff}$ side is unaffected). 
Attributing it to improper non-LTE corrections seems unlikely, because adjustment 
only at the higher $T_{\rm eff}$ side by changing the parameters of non-LTE calculations 
(e.g., collision cross section) is difficult.

As another possibility, the problem may be related to the structure of model atmosphere 
affected by convection, because the strength of such weak (i.e., deep-forming) and $T$-sensitive 
Rb~{\sc i} 7800 line is sensitive to the condition of the lower photosphere, which critically 
depends upon the treatment of convection (i.e., mixing length, overshooting) especially 
in F-type stars. The models adopted in this study are based on Kurucz's (1993a) standard ATLAS9 
model atmosphere grid computed with the mixing length of $l = 1.25 H_{p}$ ($H_{p}$: pressure 
scale height) by including the convective overshooting effect (OVERWT=1). 
Although this ``with overshooting'' model is known to represent the 
solar observational data fairly well, it is not necessarily adequate for the higher-$T_{\rm eff}$ 
F-type star Procyon (F5~IV--V), for which ``no overshooting'' model seems more suitable
(cf. Takeda et al. 1996). So, with the consideration that no-overshooting model may be more
preferable especially for F-type stars, Rb abundances were re-determined by using
no-overshooting (OVERWT=0) ATLAS9 models for all 47 Hyades stars (group 1) in order to see 
how much abundance changes [$\delta_{\rm os} \equiv A$(no overshooting)$- A$(with overshooting)] 
would result. Interestingly, as shown by red crosses in Fig.~4b, the use of no-overshooting 
models tends to decrease $A$ ($\delta_{\rm os} < 0$) at $T_{\rm eff} \gtrsim 5500$~K with the extent 
of variation ($|\delta_{\rm os}|$) increasing with $T_{\rm eff}$ (up to $\delta_{\rm os} \sim 0.04$~dex at 
$T_{\rm eff} \sim 6300$~K) (which reflects that switching-off overshooting tends to strengthen
the line because of the steeper $T$-gradient), 
while abundances are almost unaffected ($\delta_{\rm os} \sim 0$) at 
$T_{\rm eff} \lesssim 5500$~K. Accordingly, switching-off the convective overshooting
tends to act in the direction of mitigating the $T_{\rm eff}$-dependent trend (Fig.~4c)
at least qualitatively, though still insufficient in the quantitative sense.    
In any event, it would be promising to improve the modelling of convective atmosphere
(e.g., by using state-of-the-art 3D hydrodynamical simulations) toward resolving
the $T_{\rm eff}$-dependent systematic effect seen in the Rb abundances.     
 
\subsection{[Rb/Fe] vs. [Fe/H] relation}

Before discussing the behaviour of [Rb/Fe], we turn to the problem of reference abundance 
already mentioned in Sect.~4.1, where the solar abundance ($A_{\odot}^{\rm N}$) was shown to be
by $\sim$~0.1--0.2~dex higher than the solar-system abundance ($A_{\rm s.s.}$). It is worth 
pointing out that this difference may be explained by the trend seen in 
the $A^{\rm N}$ vs. $T_{\rm eff}$ relation described in Sect.~4.2 and Sect.~4.3. That is,
since the Sun is in the $T_{\rm eff}$ range ($\gtrsim 5500$~K) where Rb abundances suffer
$T_{\rm eff}$-dependent overestimation effect, $A_{\odot}^{\rm N}$ is expected to be
overestimated by $\sim 0.1$~dex (as seen from the gradient observed in Fig.~4c), 
which may be the cause for such an apparent discrepancy (despite that the true photospheric 
Rb abundance is the same as $A_{\rm s.s.}$). Accordingly, careful consideration is 
required regarding the fiducial Rb abundance adopted for evaluation of [Rb/Fe]. 

One possibility might be to use the solar abundance ($A_{\odot}^{\rm N}$) as usual, 
while limiting the sample to those of mid-G through F type (like most of the group~2 stars).   
In this case, however, while a reasonable zero point is accomplished since the centre of 
Rb abundance distribution is well represented by $A_{\odot}^{\rm N}$, a larger 
dispersion would result because abundance data of different $T_{\rm eff}$-dependent 
overestimation are mixed (cf. Fig.~9d).

Therefore, as an alternative approach, all objects at $T_{\rm eff} \gtrsim 5500$~K 
were decided not to use for [Rb/Fe] analysis in this paper. That is, 
(i) the sample stars are restricted only to those of group~3 and group~4 (i.e., dwarfs 
and giants at $T_{\rm eff} \lesssim 5500$~K), and (ii) [Rb/Fe] ratio is calculated 
by referring to $A_{\rm s.s.}$ (instead of $A_{\odot}^{\rm N}$) as [Rb/Fe] 
$\equiv A^{\rm N} - 2.36-[{\rm Fe}/{\rm H}]$, which is a reasonable choice because
$A_{\rm s.s.}$ well matches the centre of Rb abundance distribution of near-solar 
metallicity stars in group~3 and group~4 (cf. Fig.~9e and Fig.~9f).
The resulting [Rb/Fe] vs. [Fe/H] relations obtained as such are illustrated in Fig.~10, 
from which the following characteristics are observed.
\begin{itemize}
\item
A similar tendency of gradually increasing [Rb/Fe] for a decrease in [Fe/H] is observed 
for both group~3 dwarfs (Fig.~10a) and group~4 giants (Fig.~10b), though a somewhat 
larger scatter is seen for the former.
\item
A quantitative inspection of the mean $\langle$[Rb/Fe]$\rangle$ values at each metallicity 
bin in the [Fe/H] range of $-0.4 \lesssim$~[Fe/H]~$\lesssim +0.2$ (cf. Fig.~10d and Fig.~10e) 
suggests that
(1) d[Rb/Fe]/d[Fe/H] gradient is $\sim -0.4$ at $-0.4 \lesssim$~[Fe/H]~$\lesssim 0.0$, 
(2) [Rb/Fe] appears rather flat ($\sim 0$) at $0.0 \lesssim$~[Fe/H]~$\lesssim +0.2$, and
(3) this trend almost equally holds for both dwarfs and giants (though $\langle$[Rb/Fe]$\rangle$ 
for the former tends to be slightly higher than the latter by $\lesssim 0.05$~dex).   
\item
This equality of [Rb/Fe] (at the same [Fe/H]) for group~3 and group~4 stars implies 
that initial Rb abundances are retained in the stellar surface for both dwarfs
or giants, which is understandable since AGB stars are not included in group~4 (many 
are red clump giants). 
\item
Such a gentle increase in [Rb/Fe] from $\sim 0$ (at [Fe/H]~$\sim 0$) to $\sim +0.2$
(at [Fe/H]~$\sim -0.5$) is favourably (at least in the qualitative sense) compared 
with the theoretical prediction from the recent chemical evolution model (depicted by 
the solid line in Fig.~10a and Fig.~10b).
\item
In summary, the overall behaviour of the [Rb/Fe] ratios in galactic disk stars 
(dwarfs as well as giants) around the solar metallicity, gradually increasing 
with a decrease in [Fe/H], is almost consistent with the result of theoretical simulation. 
\end{itemize}

\subsection{Comparison with previous studies}

In Fig.~10c are plotted the published results of [Rb/Fe] ratios against [Fe/H],
which were reported in three previous papers (cf. Sect.~1):  Gratton \& Sneden (1994),
Tomkin \& Lambert (1999), and Abia et al. (2020).

Regarding Gratton \& Sneden (1994) and Tomkin \& Lambert (1999), only a small number 
of disk stars ($-1 \lesssim$~[Fe/H]~$\lesssim 0$) were included in their sample (since their 
interest was placed mainly on metal-poor halo stars of [Fe/H]~$\lesssim -1$)
which tend to show subsolar (negative) [Rb/Fe] ratios by a few tenths dex.
This may be due to the fact (at least partly) that they used the solar abundance 
(2.6) as the reference abundance for the zero point of [Rb/H], which is by 0.2~dex
higher than the meteoritic abundance that would have been more appropriate 
(note that most of their sample stars are at $T_{\rm eff} \lesssim 5500$~K).

In contrast, the consequence obtained in this study is in apparent conflict with 
Abia et al.'s (2020) [Rb/Fe] vs. [Fe/H] correlation derived from M dwarfs of near-solar 
metallicity, showing a controversial trend that [Rb/Fe] (being appreciably subsolar by 
several tenths dex) steeply increase towards higher metallicity within a narrow range of 
$-0.3 \lesssim$~[Fe/H]~$\lesssim +0.3$. It is hard to understand why they obtained
such a conclusion. 

It should be remarked here that their [Rb/Fe] analysis seems to be based on LTE abundances
(while using the LTE solar abundance of 2.47 as the reference) despite that 
they also derived non-LTE Rb abundances by applying (negative) non-LTE corrections
ranging from $-0.28$ to $-0.13$~dex (tending to decrease towards lower $T_{\rm eff}$; 
cf. Fig.~4 therein). Since their non-LTE correction for the Sun is $-0.12$~dex, 
non-LTE [Rb/Fe] would be shifted further downward (by $\sim 0.1$~dex on the average),
making the anomaly even more exaggerated, which may be the reason why they did not 
employ non-LTE abundances.

It seems premature, however, to regard that the validity of their non-LTE correction 
results derived for M dwarfs is established. Since the non-LTE formation 
of the Rb~{\sc i} resonance line is determined by the delicate interplay between 
two opposite effects of line weakening and line strengthening (cf. Appendix~A), 
its net result could be sensitive to detailed computational conditions.
For example, Pavlenko \& Magazz\`{u} (1996) showed in their line formation calculation
for the Li~{\sc i} 6708 line in G--M dwarfs and subgiants that the non-LTE effect
can either strengthen or weaken the line at $T_{\rm eff} \lesssim 4000$~K.
Accordingly, ``positive'' non-LTE corrections might be possible depending on 
the situation also for the case of Rb~{\sc i} line formation in M dwarfs, by which 
the apparent anomaly of subsolar [Rb/Fe] at issue could be mitigated. 

\section{Summary and conclusion}

Spectroscopic studies of stellar rubidium abundances in context of galactic chemical 
evolution have been rather scarce, presumably due to the technical difficulty of its 
abundance determination where weak and blended Rb~{\sc i} line has to be generally invoked. 

Recently, Abia et al. (2020) determined the Rb abundances of 57 M-type dwarfs of 
$-0.3 \lesssim$~[Fe/H]~$\lesssim +0.3$ and reported the anomalous trends of
[Rb/Fe] ratios, being appreciably subsolar by several tenths dex and increasing
towards higher [Fe/H], which can not be explained by any current theory
of galactic nucleosynthesis.

With an aim of examining whether such a tendency is actually observed in FGK-type stars, 
Rb abundances were determined by applying the spectrum fitting method to the 
Rb~{\sc i} 7800 line for $\sim 500$ late-type dwarfs as well as giants (including Hyades 
cluster stars) belonging to the galactic disk population ($-0.6 \lesssim$~[Fe/H]~$\lesssim +0.3$).

The Rb abundance of the Sun determined from the solar spectrum turned out to be 
by $\sim$~0.1--0.2~dex higher than the solar system abundance derived from meteorites,
as already reported in the earlier work by Gratton \& Sneden (1994) and Tomkin \& Lambert (1999).
This difference may be related to the $T_{\rm eff}$-dependent effect seen in F--G dwarfs (see below).

An inspection of the results derived for Hyades dwarfs revealed that they show 
a systematic $T_{\rm eff}$-dependent trend at $\gtrsim 5500$~K, which means that 
the Rb abundances derived for early-G to F stars (including the Sun) are likely to 
suffer some overestimation increasing towards higher $T_{\rm ef}$. This might be 
due to some imperfect treatment of surface convection in classical model atmospheres, 
because $T$-sensitive Rb~{\sc i} line is weak and forms deep in the photosphere 
where the structure is critically affected by how the convection is treated
(especially in F-type stars). 

For this reason, in the present study of [Rb/Fe], the final samople was limited to only late-G 
and K stars of $T_{\rm eff} \lesssim 5500$~K and the meteoritic abundance (instead of the 
solar abundance) was employed as the reference Rb abundance,

The resulting [Rb/Fe] vs.[Fe/H] relations for field G--K dwarfs and giants turned out to be 
similar to each other, both showing a gradual increase of [Rb/Fe] with a decrease 
in [Fe/H] (with a d[Rb/Fe]/d[Fe/H] gradient of $\sim -0.4$ around the solar metallicity).
This trend is almost consistent (at least qualitatively) with the theoretical prediction 
of chemical evolution models.

Accordingly, this study could not confirm the anomalous behaviours of 
[Rb/Fe] derived by Abia et al. (2020). The reason why they obtained such results for 
M dwarfs is yet to be clarified. 

\section*{Acknowledgments}

This research has made use of the SIMBAD database, operated by CDS, 
Strasbourg, France. 
This work has also made use of the VALD database, operated at Uppsala 
University, the Institute of Astronomy RAS in Moscow, and the University of Vienna.
This investigation is based in part on the data collected at Subaru Telescope, 
which is operated by the National Astronomical Observatory of Japan.


\clearpage
\onecolumn

\setcounter{table}{0}
\begin{table}[h]
\footnotesize
\caption{Properties of target stars.}
\begin{center}
\begin{tabular}{ccccc}\hline\hline
   Group  &         1           &        2         &            3         &             4       \\
\hline
Number of stars&        47           &       $^{*}$100        &           101        &            239      \\
  Star type   & Hyades cluster stars &    field stars   &        field stars   &   $^{\S}$field stars\\
  Sp. class   &   mid F--mid K      & early F-early K  &      late G--mid K   &       mid G--early K \\
  Lum. class  &      dwarfs         & dwarfs (+subgiants) &          dwarfs      &           giants    \\
  $^{\dagger}T_{\rm eff}$  &   [4525, 6307]      &   [5150, 6968]   &       [4495, 5482]   &    [4494, 5624] \\
  $^{\dagger}\log g$   &   [4.08, 4.71]      &   [3.38, 4.72]   &       [4.22, 4.73]   &      [1.42, 3.49] \\ 
  $^{\dagger}[$Fe/H$]$ &  [$-0.02$, +0.33]   &  [$-0.90$, +0.46]  &      [$-1.27$, +0.29]  &   [$-0.77$, +0.19] \\
  $^{\#}$Obs. data   &  TH20               &   $^{*}$T05a           &           TH20       &          T15      \\
  $^{\#}$Parameters  &  TH20               &   T05b           &           TH20       &          TSM08    \\
   Remarks    &validity check for $A$(Rb) & not used for [Rb/Fe] & adopted for [Rb/Fe]  &  adopted for [Rb/Fe] \\
\hline
\end{tabular}
\end{center}
\footnotesize
$^{\S}$Including 4 Hyades giants.\\
$^{\dagger}$Parenthesised values indicate [minimum, maximum].\\
$^{\#}$Previous papers of the author, which describe the observational data and atmospheric
parameters of each star: TH20 $\cdots$ Takeda \& Honda (2020), T05a $\cdots$ Takeda et al. (2005a),
T05b $\cdots$ Takeda et al. (2005b), T15 $\cdots$ Takeda et al. (2015),
TSM08 $\cdots$ Takeda et al. (2008).\\
$^{*}$Although the Okayama spectrum database published in T05a includes spectra of 160 FGK stars 
in total, many of those around $\sim 7800$~\AA\ are unfortunately damaged by serious bad columns 
in CCD. Only the data of 100 stars among them turned out to be somehow usable.    
\end{table}

\setcounter{table}{1}
\begin{table}[h]
\scriptsize
\caption{Adopted atomic data of Rb lines.}
\begin{center}
\begin{tabular}{cccl}\hline\hline
$\lambda$ & $\chi_{\rm low}$ & $\log gf$ & isotope \\
(\AA)  & (eV)  & (dex) &  \\
\hline
7800.183 & 0.000 & $-0.663$ & $^{87}$Rb \\
7800.186 & 0.000 & $-0.663$ & $^{87}$Rb \\
7800.188 & 0.000 & $-1.061$ & $^{87}$Rb \\
7800.317 & 0.000 & $-0.216$ & $^{87}$Rb \\
7800.322 & 0.000 & $-0.663$ & $^{87}$Rb \\
7800.325 & 0.000 & $-1.362$ & $^{87}$Rb \\
7800.233 & 0.000 & $-0.744$ & $^{85}$Rb \\
7800.234 & 0.000 & $-0.647$ & $^{85}$Rb \\
7800.235 & 0.000 & $-0.760$ & $^{85}$Rb \\
7800.292 & 0.000 & $-0.282$ & $^{85}$Rb \\
7800.295 & 0.000 & $-0.647$ & $^{85}$Rb \\
7800.296 & 0.000 & $-1.191$ & $^{85}$Rb \\
\hline
7947.507 & 0.000 & $-1.204$ & $^{87}$Rb \\
7947.524 & 0.000 & $-1.903$ & $^{87}$Rb \\
7947.651 & 0.000 & $-1.204$ & $^{87}$Rb \\
7947.668 & 0.000 & $-1.204$ & $^{87}$Rb \\
7947.563 & 0.000 & $-0.803$ & $^{85}$Rb \\
7947.570 & 0.000 & $-1.347$ & $^{85}$Rb \\
7947.626 & 0.000 & $-0.900$ & $^{85}$Rb \\
7947.634 & 0.000 & $-0.803$ & $^{85}$Rb \\
\hline
\end{tabular}
\end{center}
These data are based on Abia et al.'s (2020) Table~1.
The $gf$ values are scaled by assuming $^{85}$Rb/$^{87}$Rb = 2.43. 
Although the Rb~{\sc i} 7800 line is essentially invoked in the present study,
the data for the Rb~{\sc i}~7947 line (the other line of the resonance doublet)
are also shown because it is used in Appendix~B. 
\end{table}

\clearpage

\setcounter{figure}{0}
\begin{figure}
\begin{center}
  \includegraphics[width=10cm]{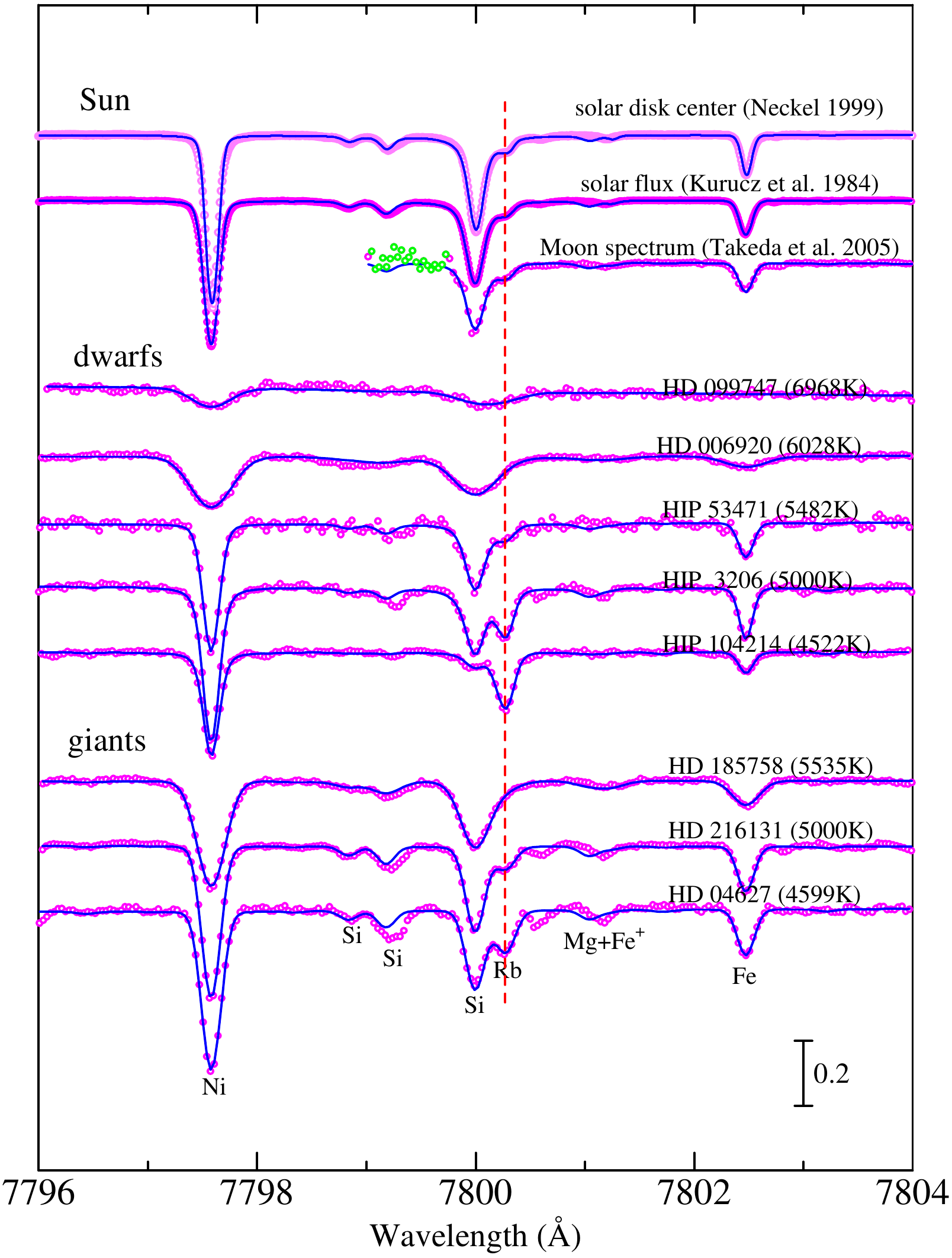}
\end{center}
\caption{
Demonstrative examples of spectrum fitting in the 7796--7804~\AA\ region 
(comprising the Rb~{\sc i} 7800 line) for selected representative cases:
the Sun (the disk centre spectrum and the flux spectrum), the Moon 
(where the fitting range had to be narrowed due to spectrum defects), 
five field dwarfs, and three field giants.
The observed and best-fit theoretical spectra are shown by open 
symbols and solid lines, respectively. Green symbols indicate the masked parts 
of observed data excluded from the fitting analysis.
The centroid wavelength of the Rb~{\sc i} line (7800.27~\AA, $gf$-weighted 
average for 12 components; cf. Table~2) is indicated by the vertical dashed line.
Each spectrum (arranged in the order of descending $T_{\rm eff}$) is vertically 
shifted by 0.2 (in unit of the continuum level) relative to the adjacent one
of the same group. The wavelength scale is adjusted to the laboratory frame 
by correcting the radial velocity shift. 
}
\label{fig:1}
\end{figure}

\setcounter{figure}{1}
\begin{figure}
\begin{center}
  \includegraphics[width=14cm]{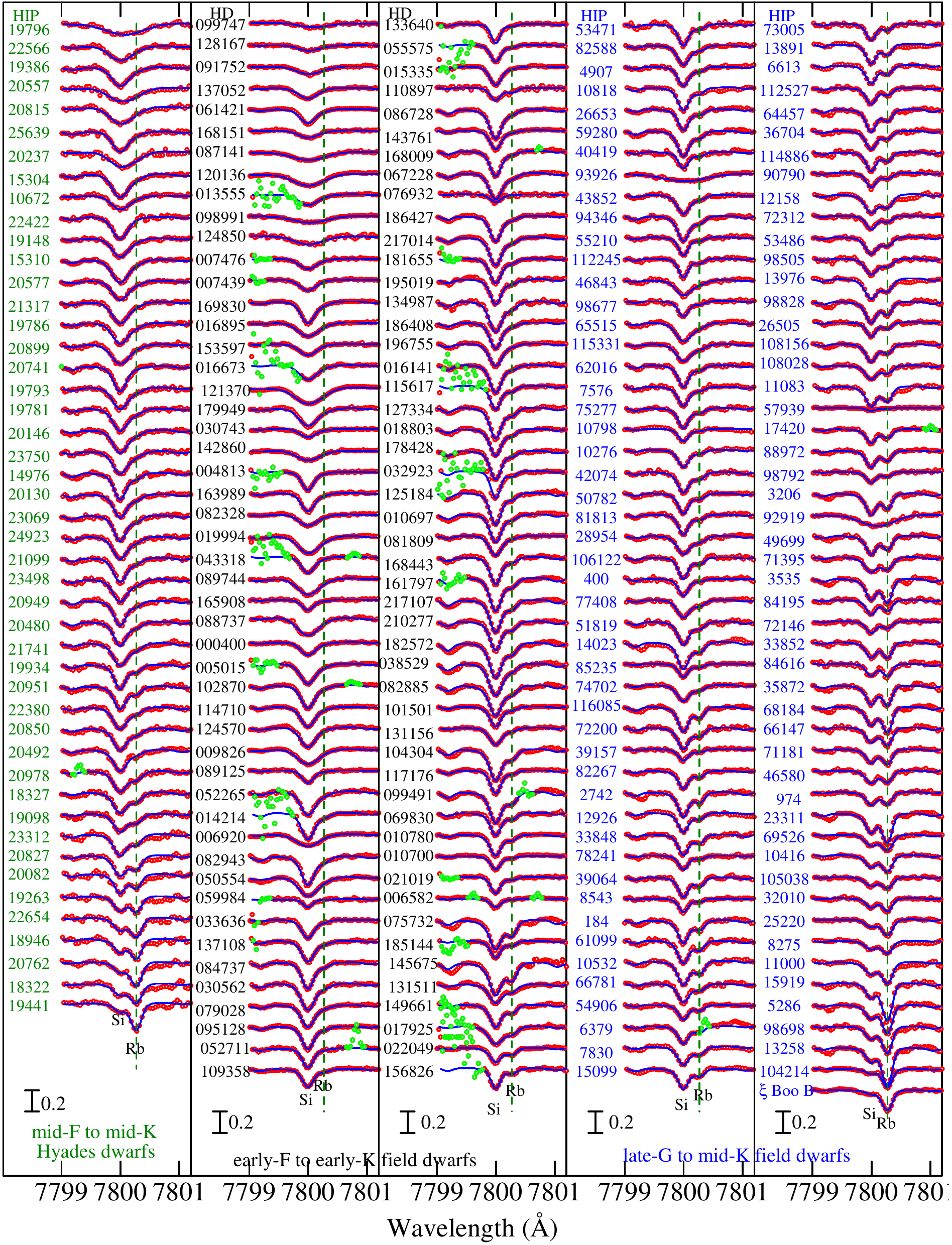}
\end{center}
\caption{
Spectrum fitting appearance in the neighbourhood  of the Rb~{\sc i} 7800 line 
(7799--7801~\AA) for 47 Hyades dwarfs (group~1; leftmost 1st panel), 
100 FGK dwarfs (group~2; 2nd and 3rd panels, where specific spectral portions had 
to be masked due to spectrum defects for not a few cases; cf. the note to Table~1), 
and 101 GK dwarfs (group~3; 4th and 5th panels). In each group, stars (indicated by 
HD or HIP numbers) are arranged in the descending order of $T_{\rm eff}$. 
Otherwise, the same as in Fig.~1.  
}
\label{fig:2}
\end{figure}

\setcounter{figure}{2}
\begin{figure}
\begin{center}
  \includegraphics[width=14cm]{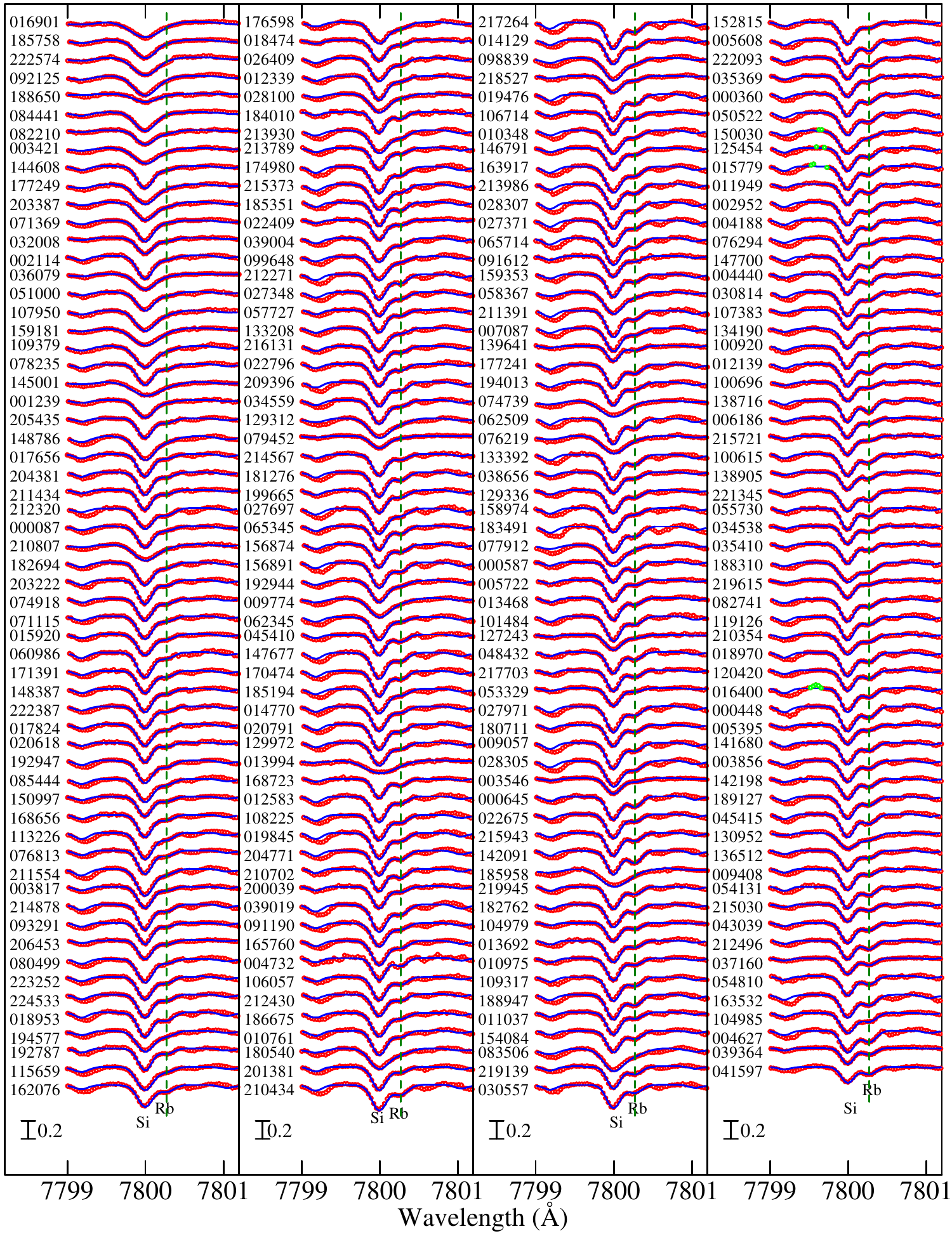}
\end{center}
\caption{
Spectrum fitting appearance in the neighbourhood of the Rb~{\sc i} 7800 line 
(7799--7801~\AA) for 239 GK giants (group~4), Stars are indicated by 
the HD numbers and arranged in the order of descending $T_{\rm eff}$. 
Otherwise, the same as in Fig.~1.
}
\label{fig:3}
\end{figure}

\setcounter{figure}{3}
\begin{figure}
\begin{center}
  \includegraphics[width=10cm]{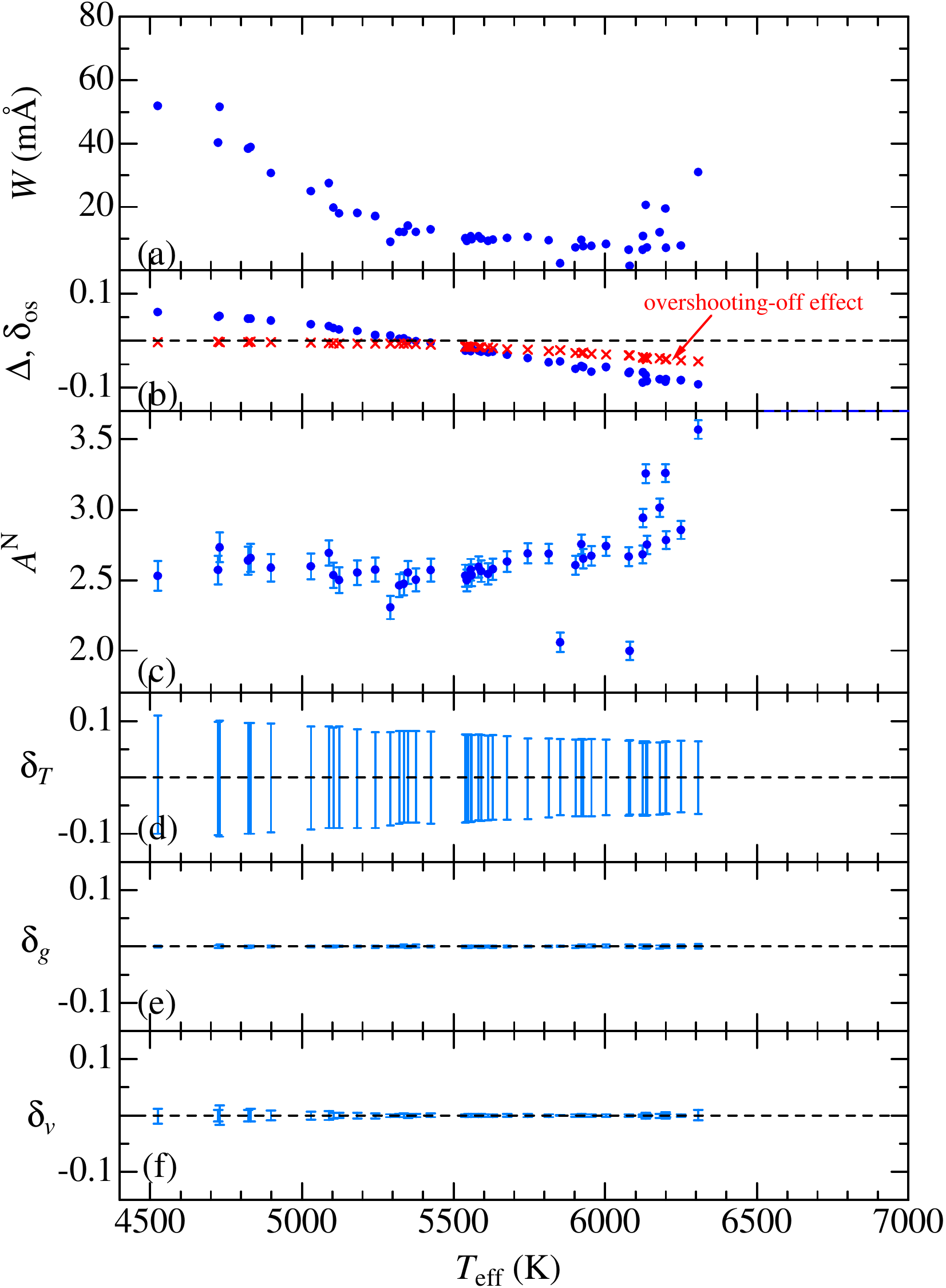}
\end{center}
\caption{
Rubidium abundances and Rb~{\sc i}~7800-related quantities derived for 
group~1 stars (47 Hyades FGK dwarfs), plotted against $T_{\rm eff}$. 
(a) $W$ (equivalent width), 
(b) $\Delta$ (non-LTE correction),
(c) $A^{\rm N}$ (non-LTE abundance)
(d) $\delta_{T+} (>0)$ and $\delta_{T-} (<0)$ (abundance variations 
in response to changing $T_{\rm eff}$ by $\pm 100$~K), 
(e) $\delta_{g+} (<0)$ and $\delta_{g-} (>0)$ (abundance variations 
in response to changing $\log g$ by $\pm 0.1$~dex), 
and (f) $\delta_{v+} (<0)$ and $\delta_{v-} (>0)$ (abundance variations in 
response to changing $v_{\rm t}$ by $\pm 0.5$~km~s$^{-1}$).
The error bars attached in the symbols in panel (c) are
$\delta_{Tgv} \equiv (\delta_{T}^{2} + \delta_{g}^{2} + \delta_{v}^{2})^{1/2}$,
where $\delta_{T} \equiv (|\delta_{T+}|+|\delta_{T-}|)/2$,
$\delta_{g} \equiv (|\delta_{g+}|+|\delta_{g-}|)/2$, and 
$\delta_{v} \equiv (|\delta_{v+}|+|\delta_{v-}|)/2$.
The red crosses plotted in panel~(b) denote the abundance changes
caused by using model atmospheres without convective overshooting 
(instead of the standard models including the overshooting effect): 
$\delta_{\rm os} \equiv A$(no overshooting)$- A$(with overshooting).
}
\label{fig:4}
\end{figure}

\setcounter{figure}{4}
\begin{figure}
\begin{center}
  \includegraphics[width=10cm]{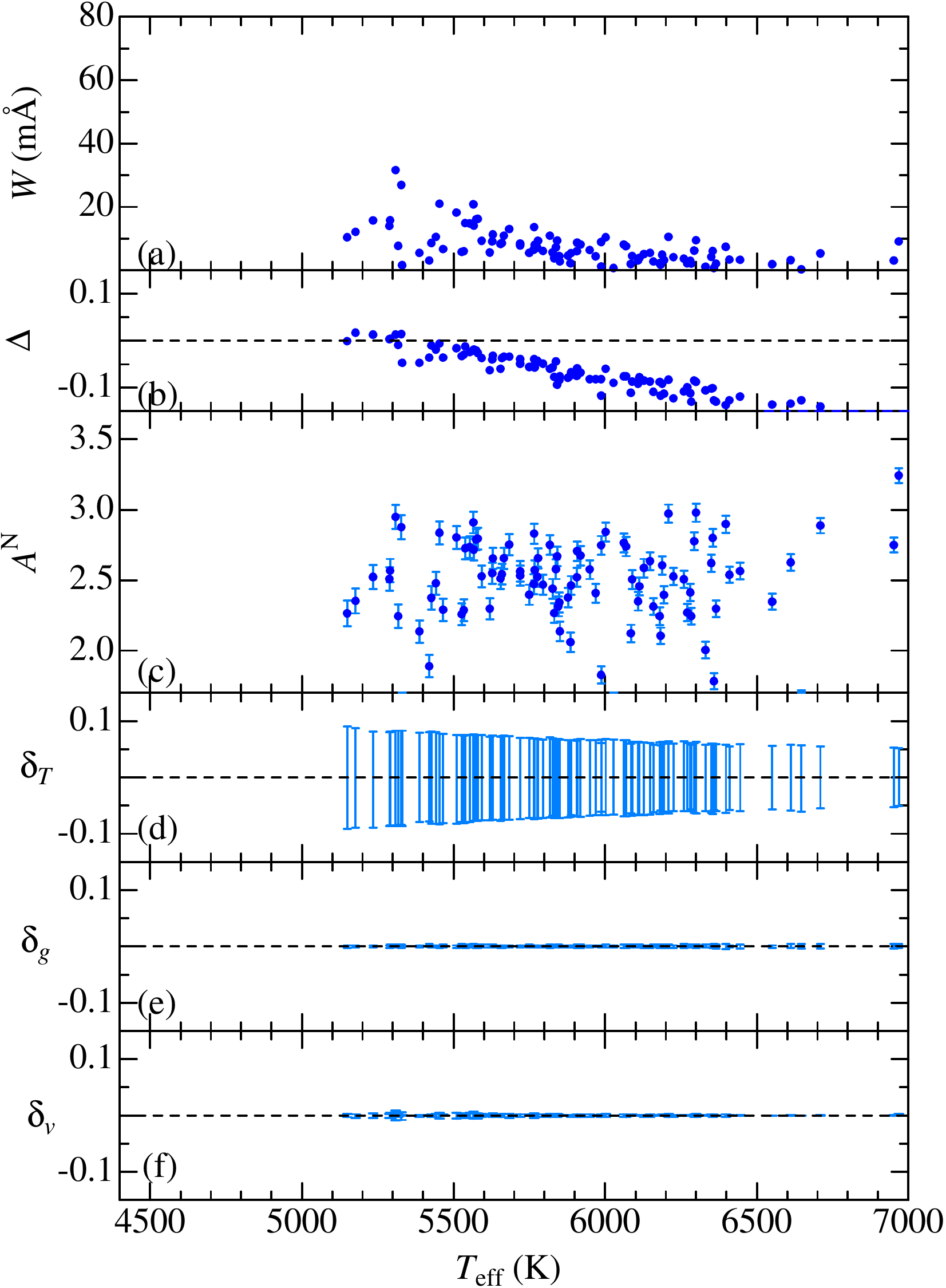}
\end{center}
\caption{
Rubidium abundances and Rb~{\sc i}~7800-related quantities 
derived for group 2 stars (100 field FGK dwarfs).
Otherwise, the same as in Fig.~4. 
}
\label{fig:5}
\end{figure}

\setcounter{figure}{5}
\begin{figure}
\begin{center}
  \includegraphics[width=10cm]{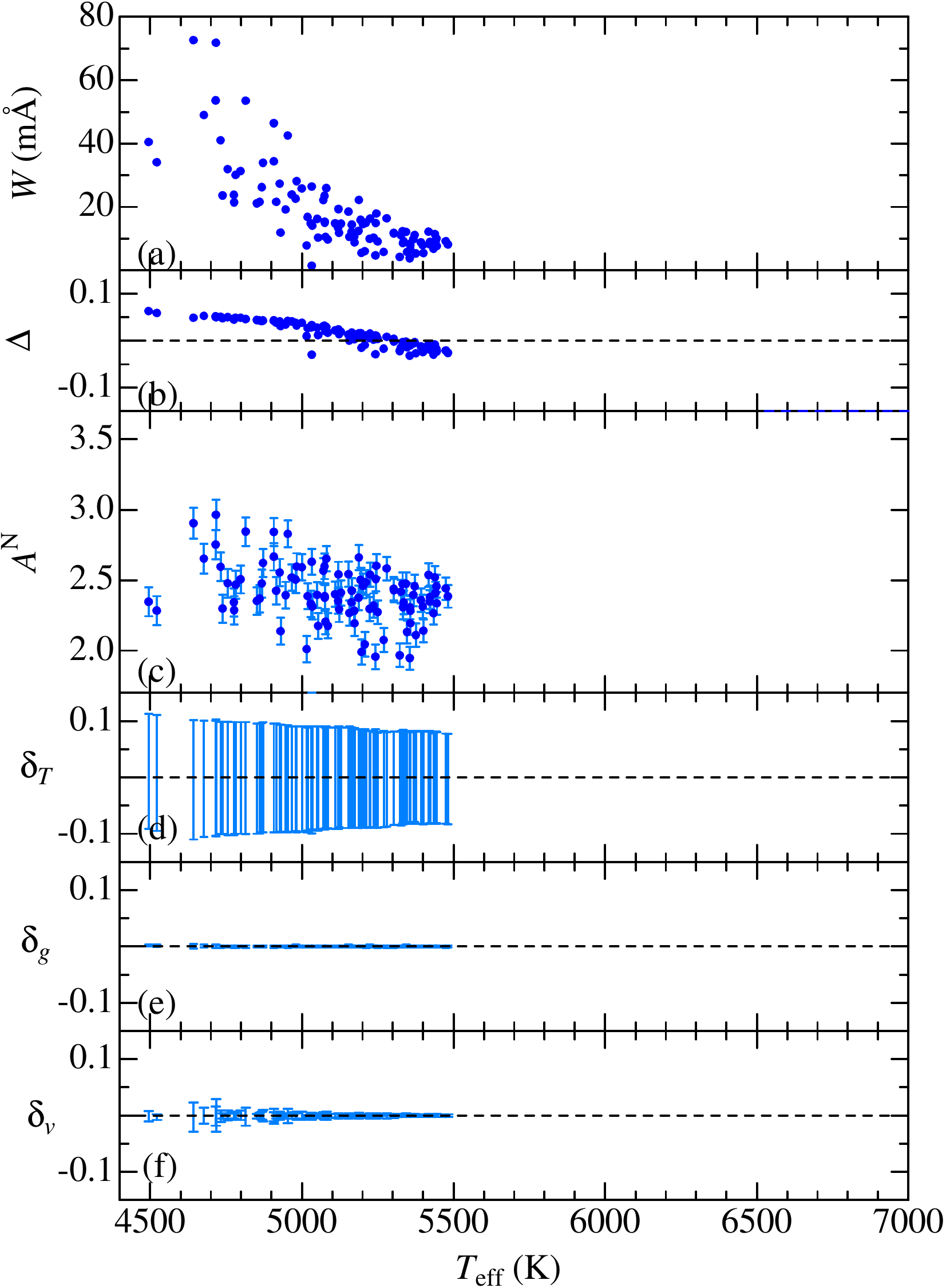}
\end{center}
\caption{
Rubidium abundances and Rb~{\sc i}~7800-related quantities 
derived for group 3 stars (101 field GK dwarfs).
Otherwise, the same as in Fig.~4. 
}
\label{fig:6}
\end{figure}

\setcounter{figure}{6}
\begin{figure}
\begin{center}
  \includegraphics[width=10cm]{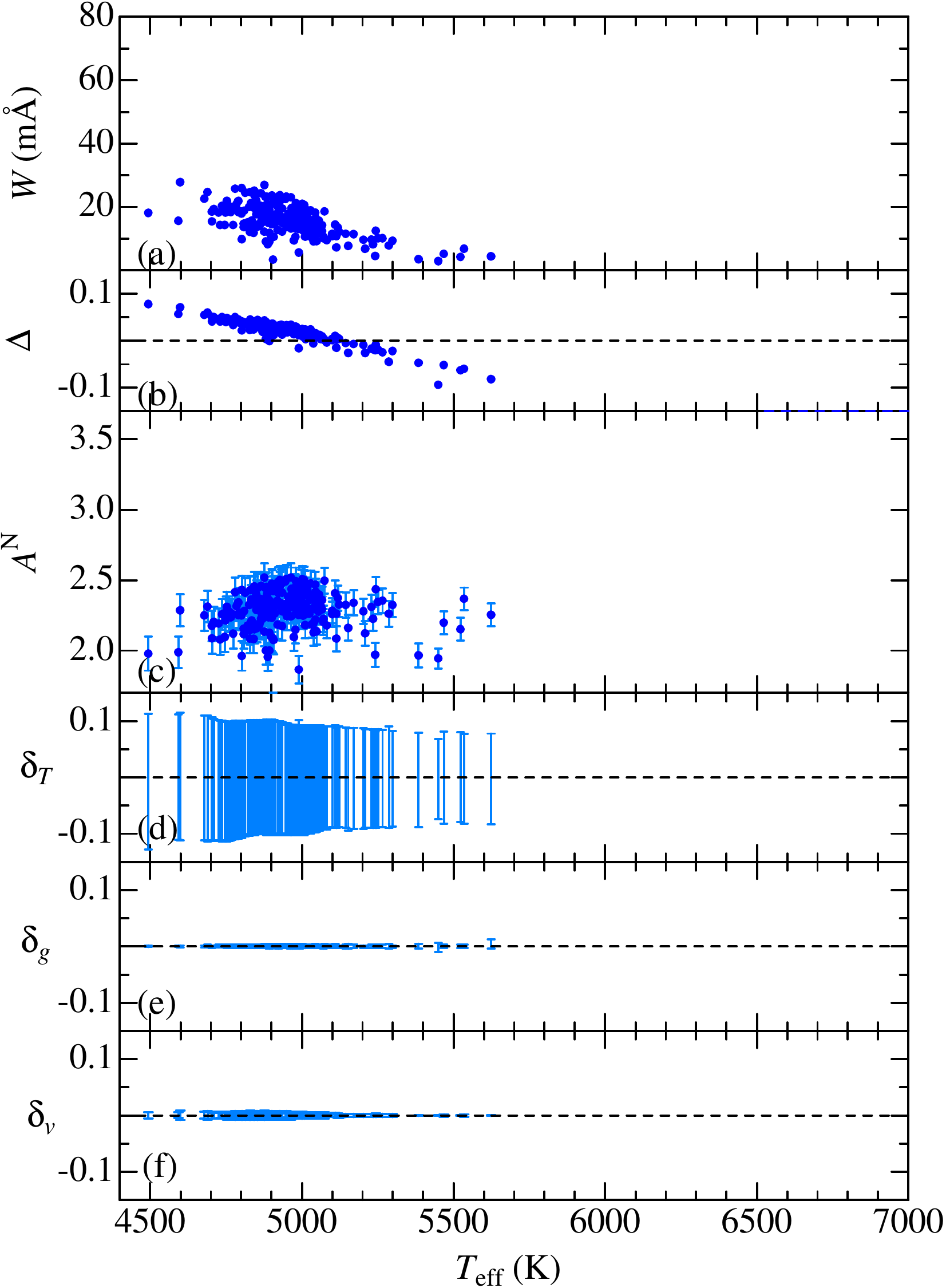}
\end{center}
\caption{
Rubidium abundances and Rb~{\sc i}~7800-related quantities 
derived for group 4 stars (239 field GK giants).
Otherwise, the same as in Fig.~4. 
}
\label{fig:7}
\end{figure}

\setcounter{figure}{7}
\begin{figure}
\begin{center}
  \includegraphics[width=6cm]{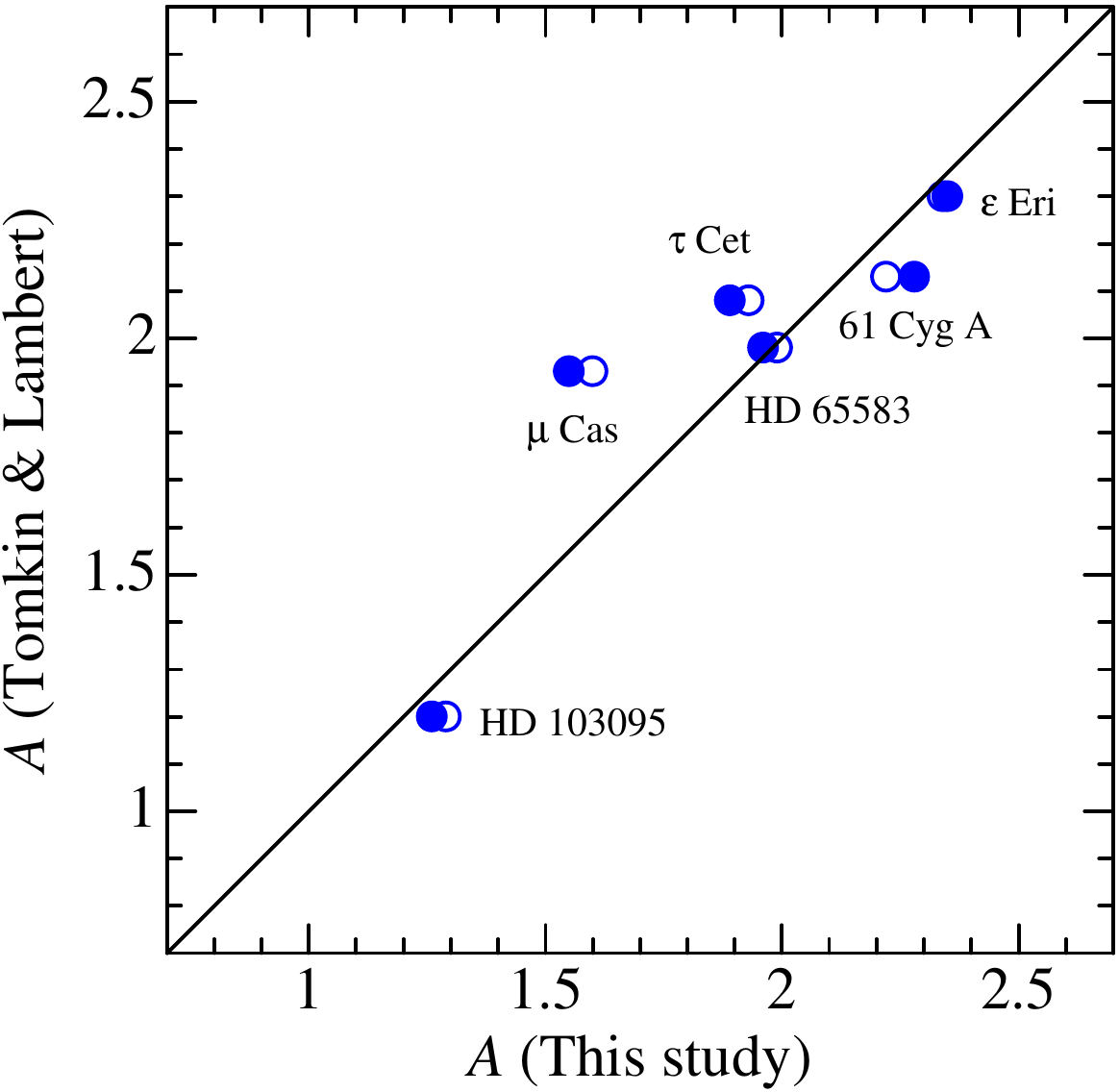}
\end{center}
\caption{
Comparison of Tomkin \& Lambert's (1999) results of Rb abundances (ordinate)  
with those derived in this study (abscissa; filled and open symbols correspond 
to non-LTE and LTE abundances, respectively) for 6 stars in common.
}
\label{fig:8}
\end{figure}

\setcounter{figure}{8}
\begin{figure}
\begin{center}
  \includegraphics[width=12cm]{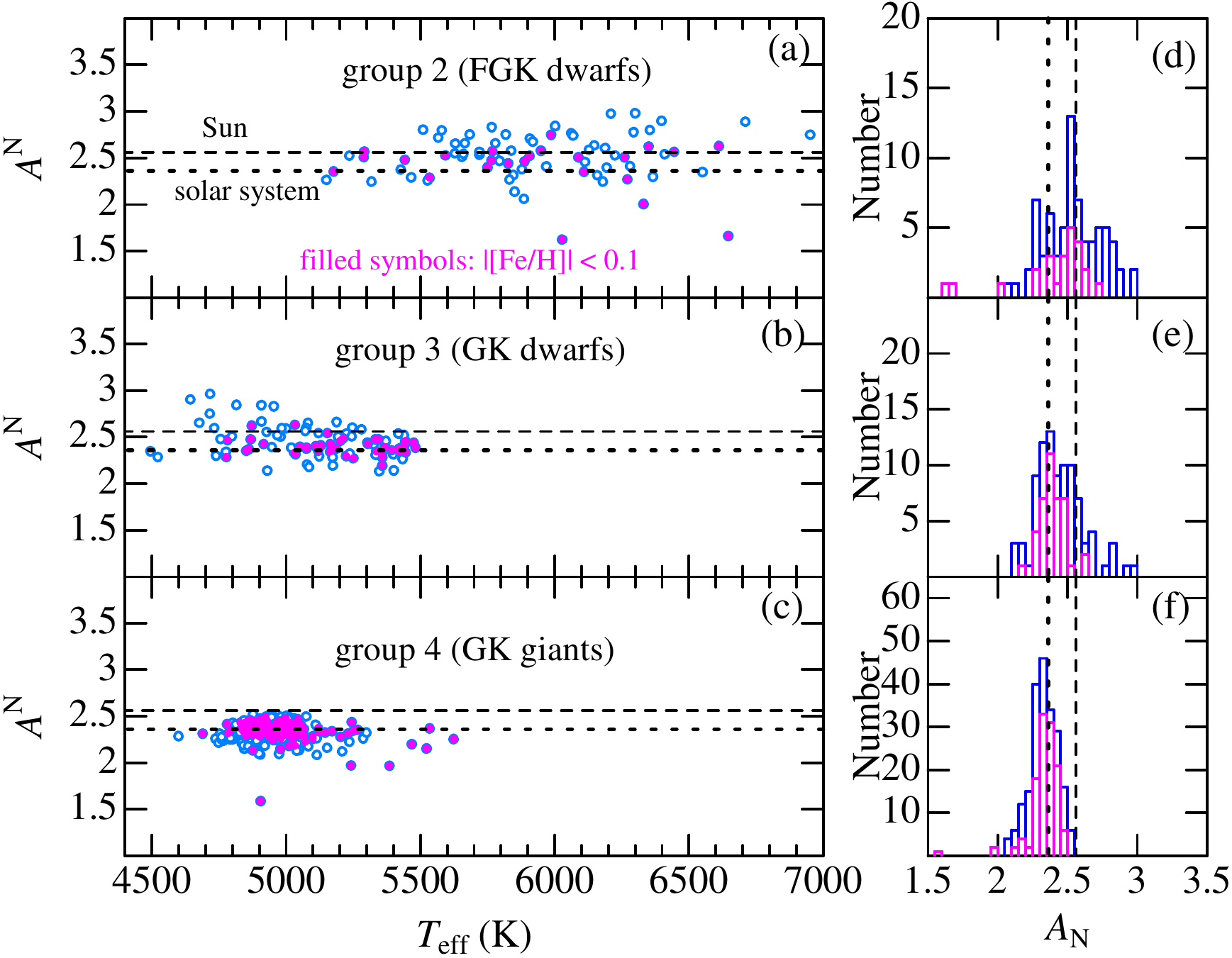}
\end{center}
\caption{
Trends of Rb abundances (left panels: $T_{\rm eff}$-dependence, right panels: 
distribution histogram) for selected near-solar metalliciy stars (sorted out 
by the criterion of $|$[Fe/H]$|<0.3$), among which those especially close to 
the solar metallicity ($|$[Fe/H]$|<0.1$) are distinguished by pink symbols
or pink lines. In the upper panels (a, d), middle panels (b, e), and lower 
panels (c, f) are shown the results for group 2, group 3, and group 4, respectively.
In each panel, the solar Rb abundance ($A_{\odot}^{\rm N} = 2.56$ derived from
the Moon spectrum) and the solar-system Rb abundance ($A_{\rm s.s.} = 2.36$) 
are indicated by dashed and dotted lines, respectively. 
}
\label{fig:9}
\end{figure}

\setcounter{figure}{9}
\begin{figure}
\begin{center}
  \includegraphics[width=12cm]{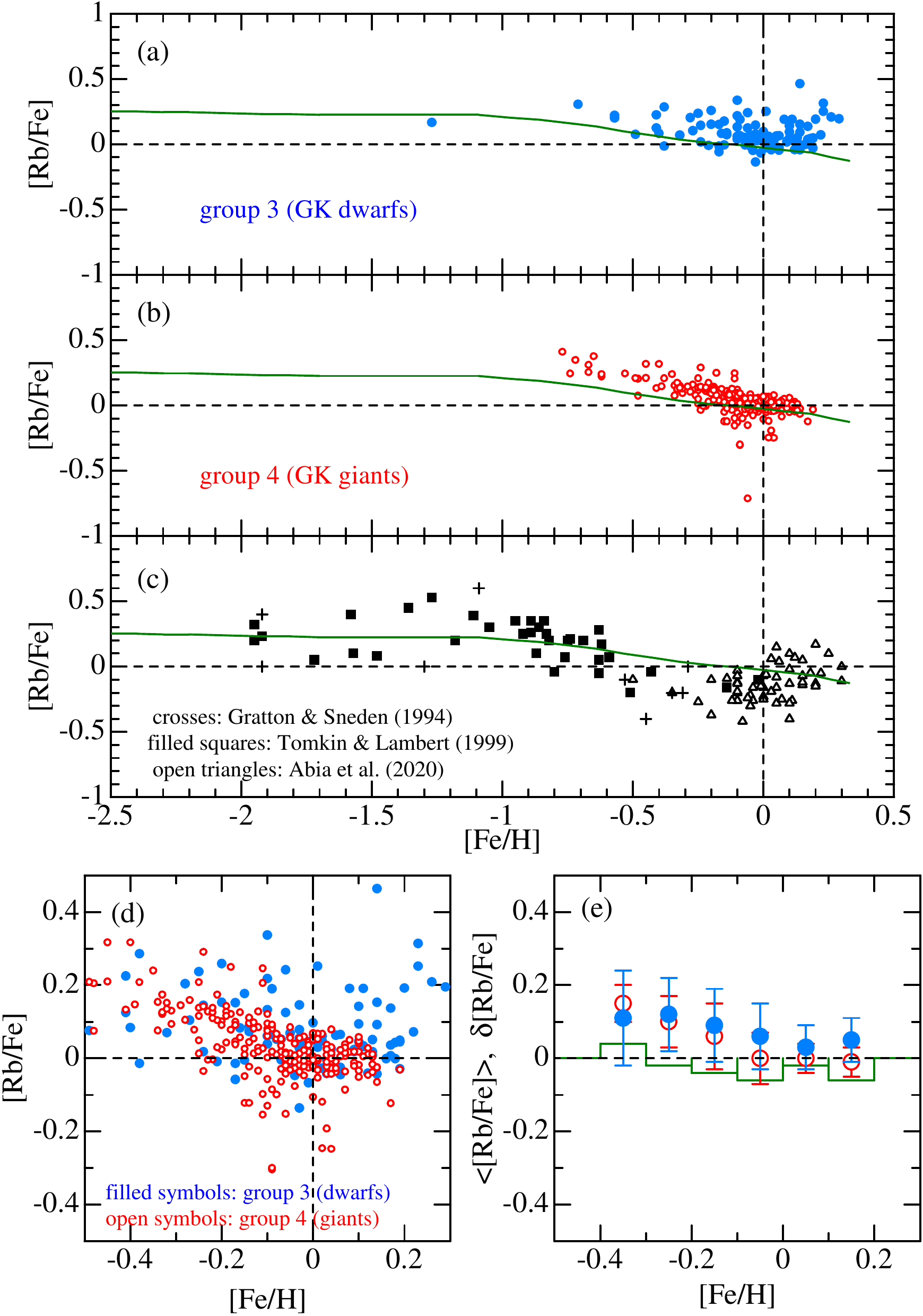}
\end{center}
\caption{
Panels (a) and (b) show the [Rb/Fe] vs. [Fe/H] correlations for group 3 (GK dwarfs)
and group 4 (GK giants) stars, respectively, where [Rb/Fe]~$\equiv A^{\rm N} - 2.36 -$[Fe/H] 
(2.36 is the solar system Rb abundance), while the published results are summarised
in panel (c) for comparison: crosses $\cdots$ Gratton \& Sneden (1994); filled squares
$\cdots$ Tomkin \& Lambert (1999); open triangles $\cdots$ Abia et al. (2020).
Note that the fiducial (solar) Rb abundances used as the zero point of [Rb/H] in these 
references are different from that adopted in this study: 2.60 (Graton \& Sneden 1994; 
Tomkin \& Lambert 1999) and 2.47 (Abia et al. 2020).
The solid line depicted in panels (a)--(c) is the theoretical [Rb/Fe] vs. [Fe/H] relation, 
which was taken from  Fig.~5 of Abia et al. (2020; case of LIMS + MSR + r-process).
The [Rb/Fe] data of groups 3 and 4 are overplotted (with the same symbols as in panels
a and b) against [Fe/H] in panel (d).
Correspondingly, large symbols in panel (e) give the mean $\langle$[Rb/Fe]$\rangle$ at 
each metallicity zone (0.1~dex bin within $-0.4 \lesssim$~[Fe/H]~$\lesssim +0.2$) 
where error bars denote the standard deviations, while bar graphs represent 
the mean abundance differences between giants (group 4) and dwarfs (group 3) defined as 
$\delta$[Rb/Fe]~$\equiv \langle$[Rb/Fe]$\rangle_{\rm giants} - \langle$[Rb/Fe]$\rangle_{\rm dwarfs}$. 
In the averaging process for deriving $\langle$[Rb/Fe]$\rangle$ at each bin, outlier data judged 
by Chauvenet's criterion were rejected.
}
\label{fig:10}
\end{figure}

\appendix

\twocolumn 

\section{Non-LTE effect on the Rb~I 7800 line}

\subsection{Physical mechanism of non-LTE line formation}

As mentioned in  Sect.~3.4, the non-LTE correction ($\Delta$) for the Rb~{\sc i} 7800 line 
monotonically increases with a decrease in $T_{\rm eff}$ from $\sim -0.1$~dex (F stars) 
to $\sim +0.1$~dex (K stars) while changing the sign at $T_{\rm eff}\sim$~5000--5500~K
(Fig.~4b, Fig.~5b, Fig.~6b, and Fig.~7b). Although this extent of $\Delta$ is quantitatively 
insignificant, the physical reason for such a $T_{\rm eff}$-dependent trend is worth explanation. 

In Fig.~A1 are shown the $l_{0}^{\rm NLTE}(\tau)/l_{0}^{\rm LTE}(\tau)$ 
(the non-LTE-to-LTE line-centre opacity ratio; almost equal to 
$\simeq b_{\rm 1}$) and $S_{\rm L}(\tau)/B(\tau)$ (the ratio of 
the line source function to the Planck function; nearly equal to 
$\simeq b_{2}/b_{1}$) for the Rb~{\sc i} 5s~$^{2}$S--5p~$^{2}$P$^{\circ}$
transition (relevant to the Rb~{\sc i} 7800 line) computed on the models of 
different $T_{\rm eff}$ and different $\log g$ while assuming [Rb/Fe]~=~[Fe/H]~=~0, 
where $b_{1}$ and $b_{2}$ are the non-LTE departure coefficients 
for the ground and first-excited terms, respectively.
Besides, the LTE and non-LTE equivalent widths ($W^{\rm L}$ and $W^{\rm N}$) 
for the Rb~{\sc i} 7800 line along with the corresponding 
non-LTE abundance correction ($\Delta$) computed for $\log g =2$ and $\log g =4$ models
are plotted against $T_{\rm eff}$ in Fig.~A2.

The upper panels of Fig.~A1 elucidate how the non-LTE effect influences
the strength of Rb~{\sc i} 7800 line for models of different $T_{\rm eff}$.
In the present case of FGK stars where Rb~{\sc i} line is comparatively weak,
two kinds of mechanisms are important both of which are caused by imbalance 
between $J$ (local mean intensity) and $B$ (Planck function at the local electron
temperature): (i) overrecombination ($J < B$) and (ii) overionisation  ($J > B$).
As we go upwards from the deep optically thick layer ($\tau > 1$, where $J=B$ holds), 
$J < B$ is first realized at the lower photosphere $\tau \lesssim 1$ (because
of the dilution of radiation due to photon escape), but eventually 
the inequality $J > B$ comes into effect at the higher optically thin region
because $J$ is almost stabilised while $B(T)$ continues to decrease
as $T$ is lowered upwards. The net non-LTE effect on the number population of the 
ground level (lower level of the Rb~{\sc i} 7800 line) is determined by
the interplay of these two mechanisms (i: line strengthening, ii: line weakening). 

In panels (a)--(c) of Fig.~A1, the action of effect (i) is seen in the overpopulation 
bump at $10^{-1} \lesssim \tau \lesssim 1$, while that of effect (ii) is manifest 
in the systematic underpopulation observed at $\tau \lesssim 10^{-2}$.
This implies that effect (i) is relatively more important at higher $T_{\rm eff}$ (where
the line is weak and deep-forming) while effect (ii) is more significant at 
lower $T_{\rm eff}$ (line is stronger and its formation region is higher).  
Moreover, this $T_{\rm eff}$-dependence of the relative importance between (i) and (ii) 
is further enhanced by the fact that the Planck function ($B$) around the ionisation edge 
becomes very $T$-sensitive as $T_{\rm eff}$ is lowered (Wien region of the Planck function)
which means that the inequality of $J > B$ causing (ii) is realized more easily.
  
What has been described above is sufficient for interpreting the non-LTE effect
on the Rb~{\sc i} 7800 line in FGK stars and its dependence upon $T_{\rm eff}$.   
Regarding  F-type stars of higher $T_{\rm eff}$ (cf. Fig.~A1c), effect (i) is more important 
and acts in the direction of strengthening the line ($\Delta < 0$).
In contrast, effect (ii) is dominant in K-type stars of lower $T_{\rm eff}$ (cf. Fig.~A1a),
where the line is weakened by the non-LTE effect ($\Delta >0$).  
For the case of G-type stars, both effects tend to compensate with each other, yielding $\Delta \sim 0$.
As such, the trends of $W^{\rm L}$, $W^{\rm N}$, and $\Delta$ with a change of $T_{\rm eff}$, 
which are displayed in Fig.~A2, can be reasonably understood. Note also that the non-LTE effect 
tends to be enhanced as the surface gravity is decreased because of the lowered atmospheric 
density (implying less collision rates).  

\subsection{Comparison with Korotin's (2020) result} 

Korotin (2020) has recently conducted non-LTE calculations for the Rb~{\sc i} 
resonance lines on an extensive model grid covering $T_{\rm eff} = $~3500--6500~K,
$\log g =$~1.0--5.0, and [Fe/H] from $-1.0$ to +0.5, which were employed
by Abia et al. (2020) for evaluating their non-LTE corrections applied to M dwarfs.  
Since his calculations also cover F-, G-, and K-type stars, it is worthwhile to compare 
the non-LTE corrections derived by both independent calculations with each other. 

By inspecting Fig.~A2b along with Korotin's (2020) Fig.~6a, the run of $\Delta$ at 
4500~$\lesssim T_{\rm eff} \lesssim 6500$~K for $\log g =2$ and $\log g =4$ 
can be compared with each other. This comparison suggests that, while the overall 
gradient (${\rm d}\Delta/{\rm d}T_{\rm eff}$) as well as its $\log g$-dependence 
is reasonably consistent, $\Delta_{\rm Korotin}$ is systematically larger than $\Delta_{\rm Takeda}$ 
by $\sim 0.1$~dex. That is, as $T_{\rm eff}$ is lowered from $\sim 6500$~K (F-type star)
to $\sim 4500$~K (K-type star), $\Delta_{\rm Takeda}$ increases from $\sim -0.1$ to $\sim +0.1$,
while $\Delta_{\rm Korotin}$ runs from $\sim -0.2$ to $\sim 0.0$.

The reason for this systematic difference can be understood by comparing the behaviour 
of $b_{1}(\tau)$ in Korotin's (2020) Fig.~3 or Fig.~7 with that of
$l_{0}^{\rm NLTE}(\tau)/l_{0}^{\rm LTE}(\tau)$ (almost equivalent to $b_{1}$)
shown in the upper panels of Fig.~A1.
As can be recognised from this comparison done for the models of same $T_{\rm eff}$, 
the overpopulation hump of $b_{1} >1$ at the lower atmosphere is apparently more conspicuous 
while the underpopulation ($b < 1$) at the upper atmosphere is less pronounced in 
the results of Korotin (2020). This means that the overionisation effect (ii) is 
repressed relative to the overrecombination effect (i) in his calculation,
which suggests that the evaluation of photoionisation rates may have been
appreciably different. Unfortunately, Korotin (2020) did not describe any details
about the calculation of photoionising radiation field ($J$), whereas
the author followed the procedure described in Sect.~3.1.2 of Takeda (1991) 
by consulting Kurucz's (1993a) ATLAS9 program along with the line opacity 
distribution function published by Kurucz (1993b).

Finally, some comments regarding the non-LTE effect in M dwarfs may be in order.
According to the calculation of Korotin (2020), $\delta (<0)$ progressively
increases as $T_{\rm eff}$ is lowered until it approaches $\sim 0$ at $T_{\rm eff} \sim 4500$~K.
However, it then shows a downturn and begins to ever decline with a decrease in $T_{\rm eff}$
over the $T_{\rm eff}$ range of $\sim$~3500--4000~K corresponding to M-type stars. 
This is presumably due to the dilution effect of line source function ($S_{\rm L}/B <1$) 
which becomes important as the line gets stronger and saturated (such as the case of 
strong K~{\sc i} resonance line at 7665/7699~\AA).
As such, the $\Delta$ values of M dwarfs calculated by Korotin (2020) remain always 
negative and this non-LTE effect (acting to strength the line) is enhanced towards 
lower $T_{\rm eff}$, as shown in Fig.~4 of Abia et al. (2020). 
Yet, it might be possible that the overionisation effect (which acts in the direction 
of line weakening) was not sufficiently taken into account in Korotin's (2020) 
calculation as mentioned above. This is the reason for having remarked the possibility 
of positive $\Delta$ in the last paragraph of Sect.~4.5. 

\onecolumn 
 
\setcounter{figure}{0}
\begin{figure}
\begin{center}
  \includegraphics[width=12cm]{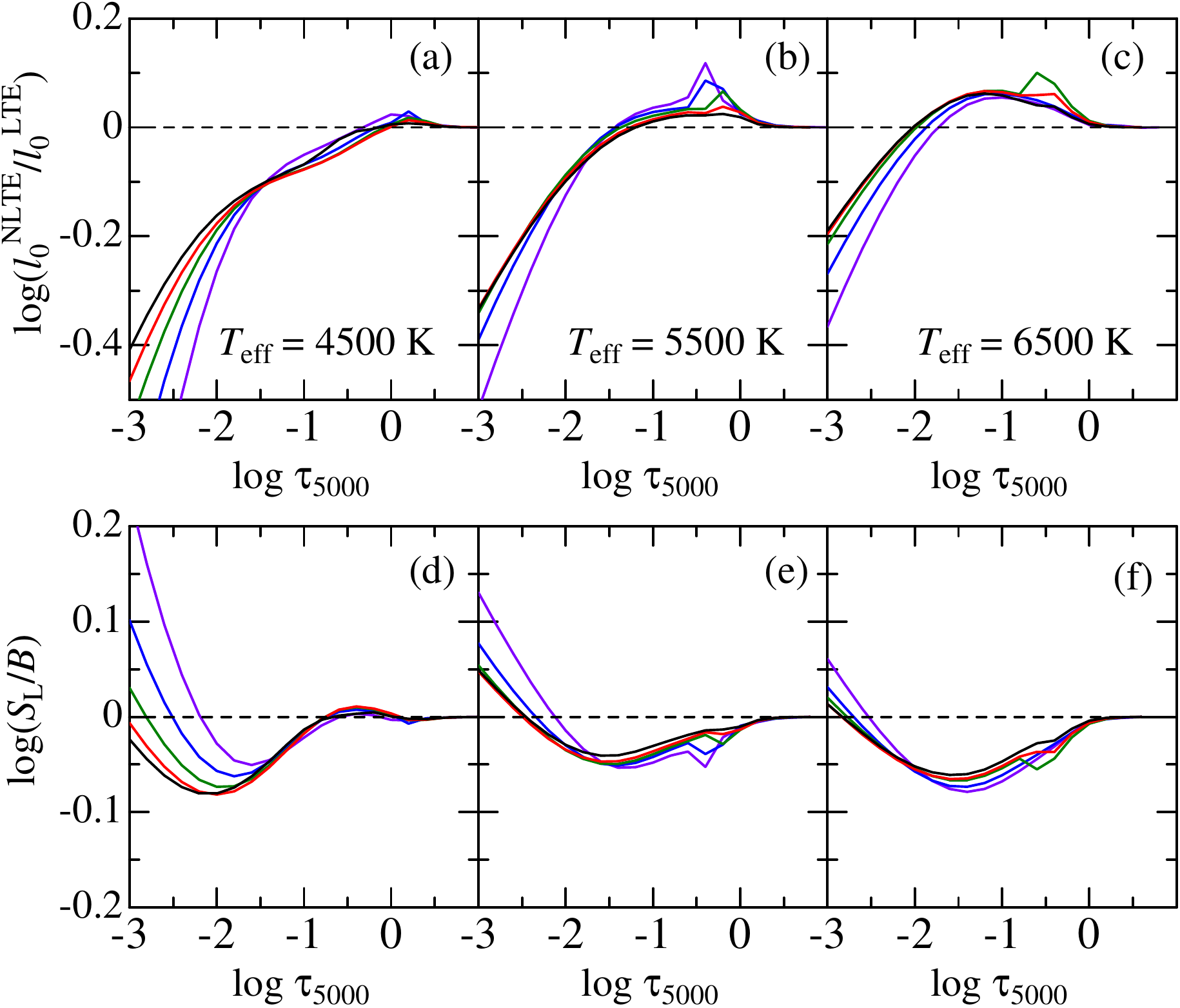}
\end{center}
\caption{
The non-LTE-to-LTE line-centre opacity ratio (upper panels a--c) and 
the ratio of the line source function ($S_{\rm L}$) 
to the local Planck function ($B$) (lower panels d--f)  
for the Rb~{\sc i} 5~$^{2}$S--5~$^{2}$P$^{\rm o}$ transition of multiplet~1, 
plotted against the continuum optical depth at 5000~\AA. 
Computations were done on the solar-metallicity models ([Fe/H] = [Rb/Fe] = 0) 
of $T_{\rm eff} =$ 4500~K (left panels a, d), 5500~K (middle panels b, e), 
and 6500~K (right panels c, f); at each panel are shown the results for 
five $\log g$ values of 1, 2, 3, 4, and 5 depicted by different colours
(violet, blue, green, red, and black, respectively). 
}
\label{fig:A1}
\end{figure}

\setcounter{figure}{1}
\begin{figure}
\begin{center}
  \includegraphics[width=9cm]{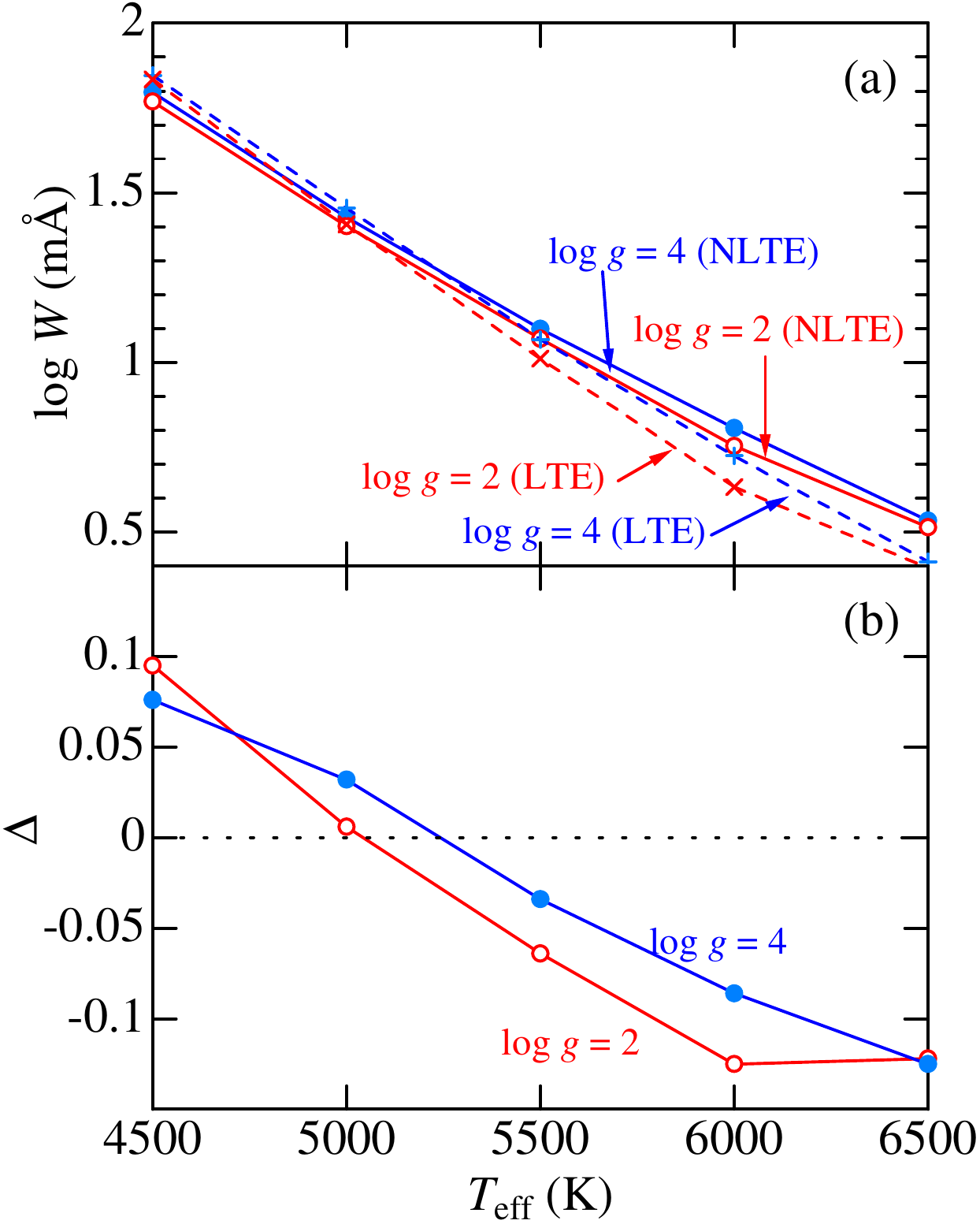}
\end{center}
\caption{
(a) Theoretical equivalent widths ($W$) of the Rb~{\sc i} 7800 line are plotted against 
$T_{\rm eff}$, where solid and dashed lines correspond to non-LTE and LTE, respectively.
(b) $T_{\rm eff}$-dependence of the non-LTE corrections ($\Delta \equiv A^{\rm N} - A^{\rm L}$). 
Shown here are the results for the case of [Rb/Fe] = [Fe/H] = 0 calculated for 
$\log g = 2$ (red) and $\log g = 4$ (blue).  
}
\label{fig:A2}
\end{figure}

\twocolumn
\section{Comparison of the abundances derived from Rb~I 7800 and 7947 lines}

As remarked in footnote~3 of Sect.~3.1, only the stronger line at 7800~\AA\ 
was invoked in this investigation, while the weaker one of the doublet at 
7947~\AA\ was not used since reliable Rb abundance determination revealed to 
be generally more difficult for this line (i.e., because of the lower quality 
of observational data along with weaker line strength). 

Grevesse et al. (2015) reported in their comprehensive analysis of solar 
photospheric abundances of Cu through Th that the Rb abundace derived from 
Rb~{\sc i} 7800 is higher than that from Rb~{\sc i}~7947 by 0.11~dex.
They thus suspected that the Rb~{\sc i} 7800 line might be blended with 
some other unknown line leading to a slight overestimation.

It is interesting to examine whether such a difference is observed in the
abundances derived from these two lines also for stars other than the Sun.
Therefore, Rb abundances were determined from the Rb~{\sc i} 7947 line
(by using the atomic data given in Table~2) for HIP~104214 (= 61~Cyg~A) 
and $\xi$~Boo~B (K5~V dwarfs with $T_{\rm eff}$ of 4523~K and 4495~K, 
respectively; lowest $T_{\rm eff}$ stars in group~3) in order to compare 
them with those derived from Rb~{\sc i} 7800 line.
Owing to sufficiently high S/N ratio (because these two stars are apparently bright) 
and comparatively large line strength (due to low $T_{\rm eff}$), Rb abundances 
from this Rb~{\sc i} 7947 line could be satisfactorily established  
by the spectrum fitting in the 7945--7050~\AA\ region (cf. Fig.~B1), 
followed by evaluations of $W$, $A^{\rm N}$, and $\Delta$ as done in Sect.~3.4.
The results obtained for two lines are presented in Table~B1.

Regarding $\xi$~Boo~B, $A^{\rm N}_{7947}$ is appreciably larger than $A^{\rm N}_{7800}$ 
by 0.15~dex (i.e., inversed inequality to that reported for the Sun).
However, this should not be seriously taken, since it must be due to influence of 
a telluric line at $\sim 7947.7$~\AA\ (see the solar spectrum shown at the top 
of Fig.~B1), which just contaminates the Rb~{\sc i} line at 7947.6~\AA\ because 
of the stellar line shift ($\sim +0.1$~\AA) corresponding to the radial velocity 
of $V_{\rm r}^{\rm topo} = +4.5$~km~s$^{-1}$ (cf. Table~B1). 

Meanwhile, Rb~{\sc i} 7947 line is essentially free from telluric blending 
for the case of HIP~104214 showing a considerably large radial velocity 
($V_{\rm r}^{\rm topo} = -82.0$~km~s$^{-1}$).
According to Table~B1, $A^{\rm N}_{7800}$ is slightly larger than $A^{\rm N}_{7947}$ 
by +0.05~dex in HIP~104214. Although this trend (sign) of abundance difference is 
similar to what Grevesse et al. (2015) reported for the Sun, its extent is so 
insignificant that the abundances from these two lines may be regarded as practically equal. 
Therefore, care should be taken in interpreting the abundance discrepancy (0.11~dex) 
between Rb~{\sc }~7800/7947 lines found by Grevesse et al. (2015).
Since the effect of telluric contamination is considerable in the solar spectrum 
(cf. Fig.~B1), the problem might be on the side of the Rb~{\sc i} 7947 line
(e.g., underestimation of the abundance due to inadequate removal of the telluric line)
rather than the blending on the Rb~{\sc i} 7800 line suggested by Grevesse et al. (2015).

\onecolumn

\setcounter{table}{0}
\begin{table}[h]
\caption{Rb abundances of two K dwarfs based on Rb~{\sc i}~7800/7947 lines.}
\begin{center}
\begin{tabular}{c r r@{  }c@{  }c r@{  }c@{  }c c l}\hline\hline
Object & $^{*}V_{\rm r}^{\rm topo}$ &$W$ & $\Delta$ & $A^{\rm N}$ &  $W$ & $\Delta$ & $A^{\rm N}$ & $\delta A^{\rm N}_{7800-7947}$ & Remark\\
    & (km~s$^{-1}$) & (m\AA)  & (dex)  & (dex) &   (m\AA)  & (dex)  & (dex)  & (dex) & \\
\hline
  &  & \multicolumn{3}{c}{(7800 line)} & \multicolumn{3}{c}{(7947 line)} & \\
HIP~104214 & $-82.0$  & 34.1 & +0.059 & 2.284  & 17.6 & +0.060 & 2.236 &  $+0.048$ & \\
$\xi$~Boo~B& $+4.5$  & 40.5 & +0.063 & 2.347  & 31.2 & +0.063 & 2.494 &  $-0.147$ & telluric contaminated (7947 line)  \\
\hline
\end{tabular}
\end{center}
$^{*}$Topocentric radial velocity (stellar radial velocity relative to
the observer). 
\end{table}

\setcounter{figure}{0}
\begin{figure}
\begin{center}
  \includegraphics[width=9cm]{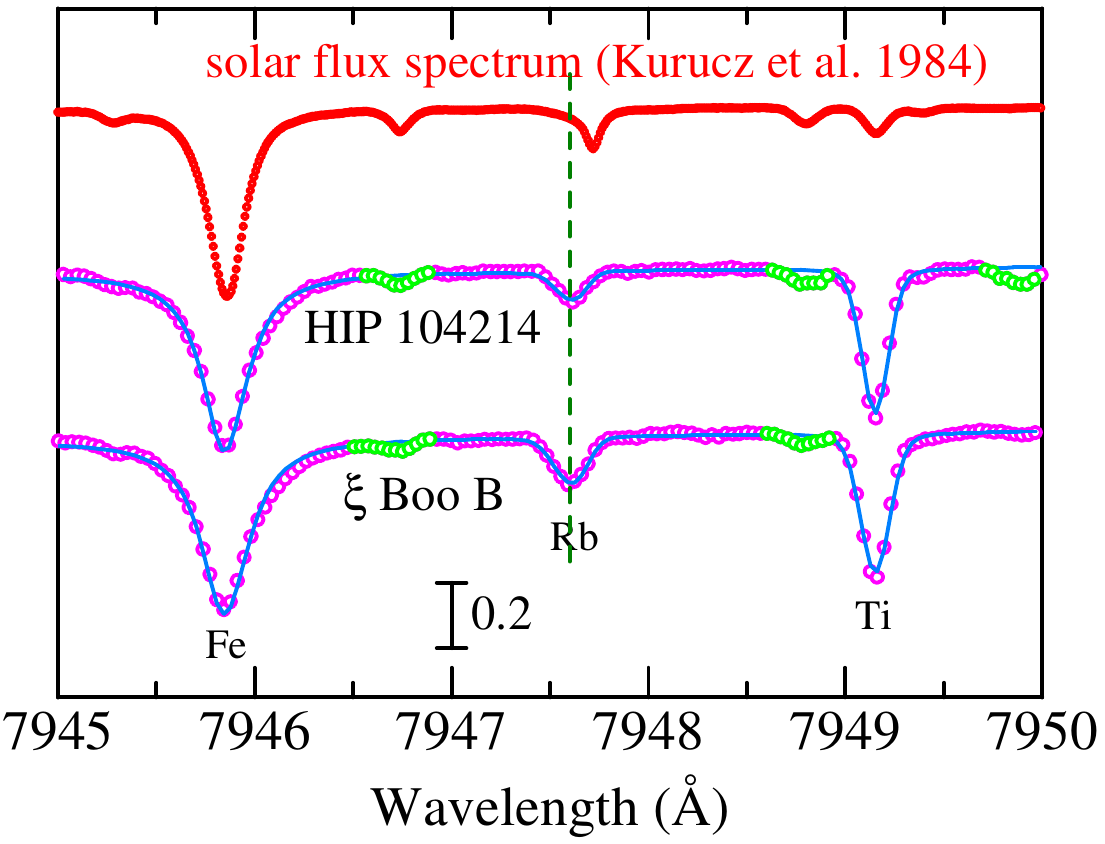}
\end{center}
\caption{
Spectrum fitting in the 7945--7950~\AA\ region comprising the Rb~{\sc i} 7947.6~\AA\ line 
(along with Fe~{\sc i} and Ti~{\sc i} lines) for HIP~104214 and $\xi$~Boo~B,
which are the lowest $T_{\rm eff}$ stars in group~3 (4522~K and 4405~K, respectively).
See the caption of Fig.~1 for the meanings of the lines and symbols.  
The centroid wavelength of the Rb~{\sc i} line (7947.60~\AA, $gf$-weighted 
average for 8 components; cf. Table~2) is indicated by the vertical dashed line.
The solar flux spectrum (Kurucz et al. 1984) is also shown at the top for comparison. 
}
\label{fig:B1}
\end{figure}


\begin{thebibliography}{body}
%
\bibitem[Abia(2020)]{Abia:2020}
  Abia, C., Tabernero, H. M., Korotin, S. A., et al., 2020, {\it A\&A}, {\it 642A}, 227. 
\bibitem[Anders(1989)]{Anders:1989} 
  Anders, E., \& Grevesse, N., 1989, {\it Geochim. Cosmochim. Acta.}, {\it 53}, 197.
\bibitem[Cayrel(1988)]{Cayrel:1988}
  Cayrel, R., 1988, in {\it The Impact of Very High S/N Spectroscopy on Stellar Physics,
  Proc. IAU Symp. 132}, eds. G. Cayrel de Strobel \& M. Spite (Dordrecht: Kluwer), p.345.
\bibitem[D'Orazi(2013)]{D'Orazi:2013}
  D'Orazi, V., Lugaro, M., Campbell, S. W., et al., 2013, {\it ApJ}, {\it 776}, 59.
\bibitem[GarciaHernandez(2006)]{GarciaHernandez:2006}
  Garc\'{\i}a-Hern\'{a}ndez, D. A., Garc\'{\i}a-Lario, P., Plez, B., D'Antona, F., 
  Manchado, A., \& Trigo-Rodr\'{\i}guez, J. M., 2006, {\it Science}, {\it 314}, 1751. 
\bibitem[Gratton(1994)]{Gratton:1994}
  Gratton, R. G., \& Sneden, C., 1994, {\it A\&A}, {\it 287}, 927. 
\bibitem[Grevesse(2015)]{Grevesse:2015}
  Grevesse, N., Scott, P., Asplund, M., Sauval, A. J., 2015, {\it A\&A}, {\it 573}, A27.
\bibitem[Korotin(2020)]{Korotin:2020}
  Korotin, S. A., 2020, {\it Astron. Lett.}, {\it 46}, 541.
\bibitem[Kurucz(1993a)]{Kurucz:1993a}
  Kurucz, R. L., 1993a, {\it Kurucz CD-ROM}, {\it No. 13} (Harvard-Smithsonian Center
  for Astrophysics).
\bibitem[Kurucz(1993b)]{Kurucz:1993b}
  Kurucz, R. L., 1993b, {\it Kurucz CD-ROM}, {\it No. 14} (Harvard-Smithsonian Center
  for Astrophysics).
\bibitem[Kurucz(1995)]{Kurucz:1995}
  Kurucz, R. L., \& Bell, B., 1995, {\it Kurucz CD-ROM}, {\it No. 23} 
  (Harvard-Smithsonian Center for Astrophysics).
\bibitem[Kurucz(1984)]{Kurucz:1984}
  Kurucz, R. L., Furenlid, I., Brault, J., \& Testerman, L., 1984,  
  {\it Solar Flux Atlas from 296 to 1300~nm}
  (Sunspot, New Mexico: National Solar Observatory).
\bibitem[Lodders(2020)]{Lodders:2020}
  Lodders, K., 2020, {\it Solar Elemental Abundances}, in
  {\it The Oxford Research Encyclopedia of Planetary Science} (Oxford University Press)
   [{\it arXiv: 1912.00844}].
\bibitem[Lowell(2002)]{Lowell:2002}
  Lowell, J. R., Northup, T., Patterson, B. M., Takekoshi, T., \& Knize, R. J., 2002,
  {\it Phys. Rev. A}, {\it 66}, 062704.
\bibitem[Nadeem(2011)]{Nadeem:2011}
  Nadeem, A., \& Haq, S. U., 2011, {\it Phys. Rev. A}, {\it 83}, 063404.
\bibitem[Neckel(1994)]{Neckel:1994}
  Neckel, H., 1994, 
  in {\it The Sun as a Variable Star, Solar and Stellar Irradiance Variations,
  IAU Coll. 143}, ed. J. M. Pap, C. Frolich, H. S. Hudson, \& S. Solanki
  (Cambridge University Press: Cambridge), p.37. 
\bibitem[Neckel(1999)]{Neckel:1999}
  Neckel, H., 1999, {\it Solar Phys.}. {\it 184}, 421.
\bibitem[Pavlenko(1996)]{Pavlenko:1996}
  Pavlenko, Ya. V., \& Magazz\`{u}, A., 1996, {\it A\&A}, {\it 311}, 961.
\bibitem[Ryabchikova(2015)]{Ryabchikova:2015}
  Ryabchikova, T., Piskunov, N., Kurucz, R. L., Stempels, H. C., Heiter, U., 
  Pakhomov, Yu, \& Barklem, P. S., 2015, {\it Phys. Scr.}, {\it 90}, 054005. 
\bibitem[Shejeelammal(2020)]{Shejeelammal:2020}
  Shejeelammal, J., Goswami, A., Goswami, P. P., Rathour, R. S., \& Masseron, T., 
  2020, {\it MNRAS}, {492}, 3708.
\bibitem[Steenbock(1984)]{Steenbock:1984}
  Steenbock, W., \& Holweger, H., 1984, {\it A\&A}, {\it 130}, 319.
\bibitem[Takeda(1991)]{Takeda:1991}
  Takeda, Y., 1991, {\it A\&A}, {\it 242}, 455.
\bibitem[Takeda(1995)]{Takeda:1995}
  Takeda, Y., 1995, {\it PASJ}, {\it 47}, 287.
\bibitem[Takeda(2020)]{Takeda:2020}
  Takeda, Y., \& Honda, S., 2020, {\it AJ}, {\it 159}, 174.
\bibitem[Takeda(1996)]{Takeda:1996}
  Takeda, Y., Kato, K., Watanabe, Y., \& Sadakane, K., 1996, {\it PASJ}, {\it 48}, 511.
\bibitem[Takeda(2005)]{Takeda:2005}
  Takeda, Y., \& Kawanomoto, S., 2005, {\it PASJ}, {\it 57}, 45.
\bibitem[Takeda(2005b)]{Takeda:2005b}
  Takeda, Y., Ohkubo, M., Sato, B., Kambe, E., \& Sadakane, K., 2005b, {\it PASJ}, {\it 57}, 27.
\bibitem[Takeda(2005a)]{Takeda:2005a}
  Takeda, Y., Sato, B., Kambe, E., et al., 2005a, {\it PASJ}, {\it 57}, 13.
\bibitem[Takeda(2008)]{Takeda:2008}
  Takeda, Y., Sato, B., \& Murata, D., 2008, {\it PASJ}, {\it 60}, 781.
\bibitem[Takeda(2015)]{Takeda:2015}
  Takeda, Y., Sato, B., Omiya, M., \& Harakawa, H. 2015, {\it PASJ}, {\it 67}, 24.
\bibitem[Takeda(2002)]{Takeda:2002}
  Takeda, Y., Zhao, G., Chen, Y.-Q., Qiu, H.-M., \& Takada-Hidai, M., 2002, {\it PASJ}, {\it 54}, 275.
\bibitem[Takeda(2003)]{Takeda:2003}
  Takeda, Y., Zhao, G., Takada-Hidai, M., Chen, Y.-Q., Saito, Y., \& Zhang, H.-W., 2003, 
  {\it ChJAA}, {\it 3}, 316.
\bibitem[Tomkin(1999)]{Tomkin:1999}
  Tomkin, J., \& Lambert, D. L., 1999, {\it ApJ}, {\it 523}, 234.
\bibitem[Yong(2006)]{Yong:2006}
  Yong, D., Aoki, W., Lambert, D. L., \& Paulson, D. B., 2006, {\it ApJ}, {\it 639}, 918.
\bibitem[Yong(2008)]{Yong:2008}
  Yong, D., Lambert, D. L., Paulson, D. B., \& Carney, B. W. 2008, {\it ApJ}, {\it 673}, 854.
\end{thebibliography}
\end{document}